\documentclass[pre,aps,twocolumn,showpacs,amsmath,amssymb,amsfonts,floatfix,superscriptaddress]{revtex4}
\usepackage{graphicx}
\usepackage{dcolumn}
\usepackage{bm}
\usepackage[hypertex]{hyperref}
\usepackage{latexsym}
\usepackage{float}
\usepackage{supertabular}
\usepackage{longtable}

\begin{document}
\title{Nonlinear $q$-voter model with inflexible zealots}
\author{Mauro Mobilia}
\affiliation{Department of Applied Mathematics, School of Mathematics, University of Leeds, Leeds LS2 9JT, U.K.} %
\email{M.Mobilia@leeds.ac.uk} %
\begin{abstract}
We study the dynamics of the nonlinear $q$-voter model with inflexible zealots in a finite  well-mixed population. In this system,  each individual supports  one of two parties and is either a susceptible voter or an inflexible zealot. At each time step, a susceptible  adopts the opinion of a neighbor if this belongs to a group of $q\geq 2$ neighbors all in the same state,
whereas inflexible zealots never change their opinion. In the presence of zealots of both parties
the model is characterized by a fluctuating stationary state and,
below a zealotry density threshold, the  
distribution of opinions is bimodal. After a characteristic time, most susceptibles become supporters
of the party having more zealots and the opinion distribution is asymmetric.
When the number of zealots of both parties is the same,
the opinion distribution is symmetric and, in the long run,  susceptibles endlessly swing from the state where they all support one party to the opposite state. Above the zealotry density threshold,  when there is an unequal number of zealots of each type, the probability distribution is single-peaked and non-Gaussian. These properties are investigated analytically and with stochastic simulations. We also study the mean time to reach a consensus when  zealots support only one party. 
\end{abstract}

\maketitle

\section{Introduction}
The voter model (VM)~\cite{Liggett} is one of the simplest and most influential examples
of individual-based systems exhibiting collective behavior. The VM has been used as a paradigm 
for the dynamics of opinion in socially interacting populations, see {\it e.g.}~\cite{Opinions,Sociophysics} and references therein. The classical, or linear, VM is closely related  to the Ising model~\cite{Glauber} and describes how consensus results from the interactions between neighboring agents endowed with a discrete set of 
states (``opinions'').
While the VM is one of the rare exactly solvable models in non-equilibrium statistical physics,
it relies on oversimplified assumptions such as perfect conformity
and lack of self-confidence  of all voters. 
This is clearly unrealistic as it is recognized that members of a society respond differently to stimuli: Many exhibit conformity while some show independence, and this influences the underlying social dynamics~\cite{Granovetter,ConfIndep,GroupSize}. In order  to  mimic the dynamics of socially interacting  agents with different levels of confidence, this author  introduced ``zealots'' in the VM~\cite{MM1,MM2,zealot07}. Originally zealots were  agents favoring one opinion~\cite{MM1,MM2}. The case of inflexible zealots whose state never changes  was then also studied~\cite{zealot07}, and the influence of committed and/or independent individuals  was considered in various models of opinion and social dynamics~\cite{MM3,otherZealots}. Recently, authors have investigated the effect of zealots in naming and cooperation games, and  even in theoretical ecology~\cite{gamesZealots,MM3}.

In recent years, many versions of the VM have been proposed~\cite{Opinions}. A particularly interesting 
variant of the VM is the two-state nonlinear $q$-voter model ($q$VM) introduced in \cite{qVM}. In this model $q$ randomly picked  neighbors may influence a voter to change its opinion. When $q=2$, 
the {\it q}VM is closely related to the  Sznajd model~\cite{Sznajd,Slanina,genSznajd} and to that of Ref.~\cite{vacillating}. The properties of the $q$VM have received much attention and there is a debate on the  expression of the exit probability in one dimension~\cite{vacillating,exitprobq2,exitprobq}.

Here, we investigate a generalization of the nonlinear $q$VM, with $q\geq 2$, in which a well-mixed population consists of inflexible zealots and  susceptible voters influenced by their neighbors. 
As a motivation, this parsimonious model allows 
to capture three important concepts of social psychology~\cite{ConfIndep} and sociology~\cite{Granovetter}: (i) conformity/imitation is an important social mechanism for collective actions; (ii)  group pressure is known to influence the degree of conformity, especially when a group size threshold is reached~\cite{GroupSize}; (iii) the degree of  conformity can be radically altered by the presence of 
some individuals that are capable of resisting group pressure~\cite{GroupSize,ConfIndep}. Here, the $q$VM mimics the process of conformity by imitation with group-size threshold, whereas zealots are independent agents that resist social pressure and can thus prevent to  reach unanimity.

In this work, we study the fluctuation-driven dynamics of the two-state $q$VM with zealots
in finite well-mixed populations and shed light on the deviations from the mean field description
and from the linear case ($q=1$). We find that below a zealotry density threshold 
the probability distribution  is bimodal instead of Gaussian and, after a characteristic time, most susceptibles  become supporters of the party having more zealots.
When both parties have the same small number of zealots, susceptibles endlessly swing from the state where they all support one party to the other with a mean switching time that approximately grows exponentially with the population size.

In the next section we introduce the model. Sections III and IV are dedicated to the mean field description  and to the model's stationary probability distribution. In Secs.~V and VI we discuss the long-time dynamics and the mean consensus time when there is one type of zealots. We summarize our findings and conclude in Sec.~VII.

\section{The $q$-voter model with zealots}
We consider a population of $N$ voters that can support one of two parties, either A or B, and therefore be in two states. Supporters of party A are in state $+1$, and those supporting party B are in state $-1$. Among the voters, a fixed number of them are ``inflexible zealots'' while the others 
are ``susceptibles''. Here, zealots are individuals that never change opinion: they permanently support either party A (A-zealots)
or party B (B-zealots). Susceptible voters can change their opinion under the pressure of a group of neighbors. The population thus consists of a number $Z_+$ of A-zealots (pinned in state $+1$) and 
$Z_-$ of B-zealots (pinned in state $-1$), and  a total of  $S=N-Z_{+}-Z_{-}$ susceptibles agents, of which $n$ are A-susceptibles (non-zealot voters in state $+1$) and $S-n$ are B-susceptibles (non-zealot voters in state $-1$). The fraction, or density, of susceptibles in the entire population remains constant and is given by $s=S/N$. For simplicity we assume that all agents have the same persuasion strength.

 At each time step, a susceptible voter consults a group of $q$
neighbors (with $q>1$) and, if there is consensus in the group, the voter is persuaded to adopt the group's state  with rate $1$~\cite{qVM}.   
The dynamics is  a generalization of the nonlinear $q$VM~\cite{qVM}  with a finite density of zealots~\cite{zealot07}, and consists of the following steps:
\begin{figure}
\includegraphics[width=3.1in, height=1.65in,clip=]{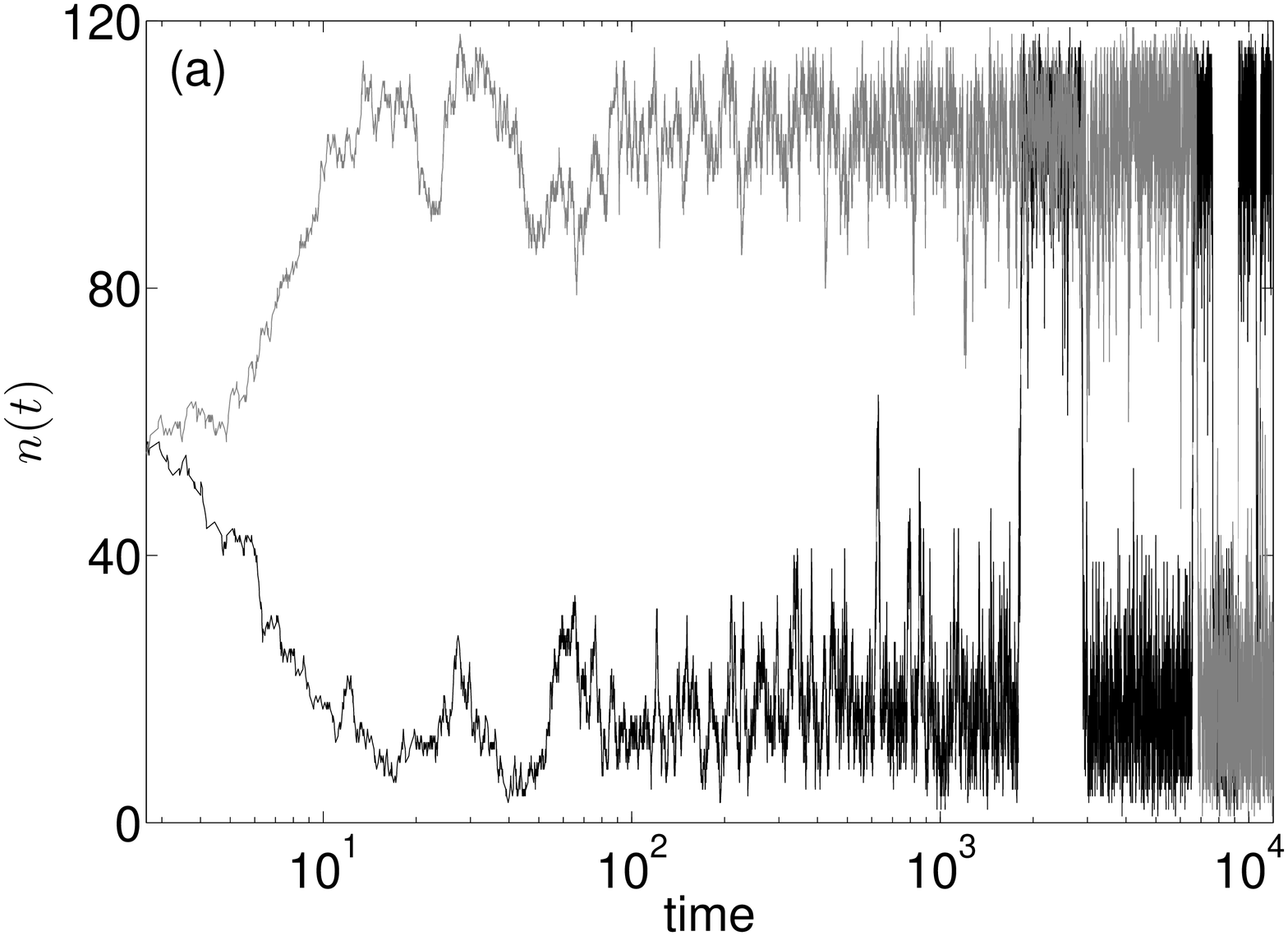}
\includegraphics[width=3.1in, height=1.65in,clip=]{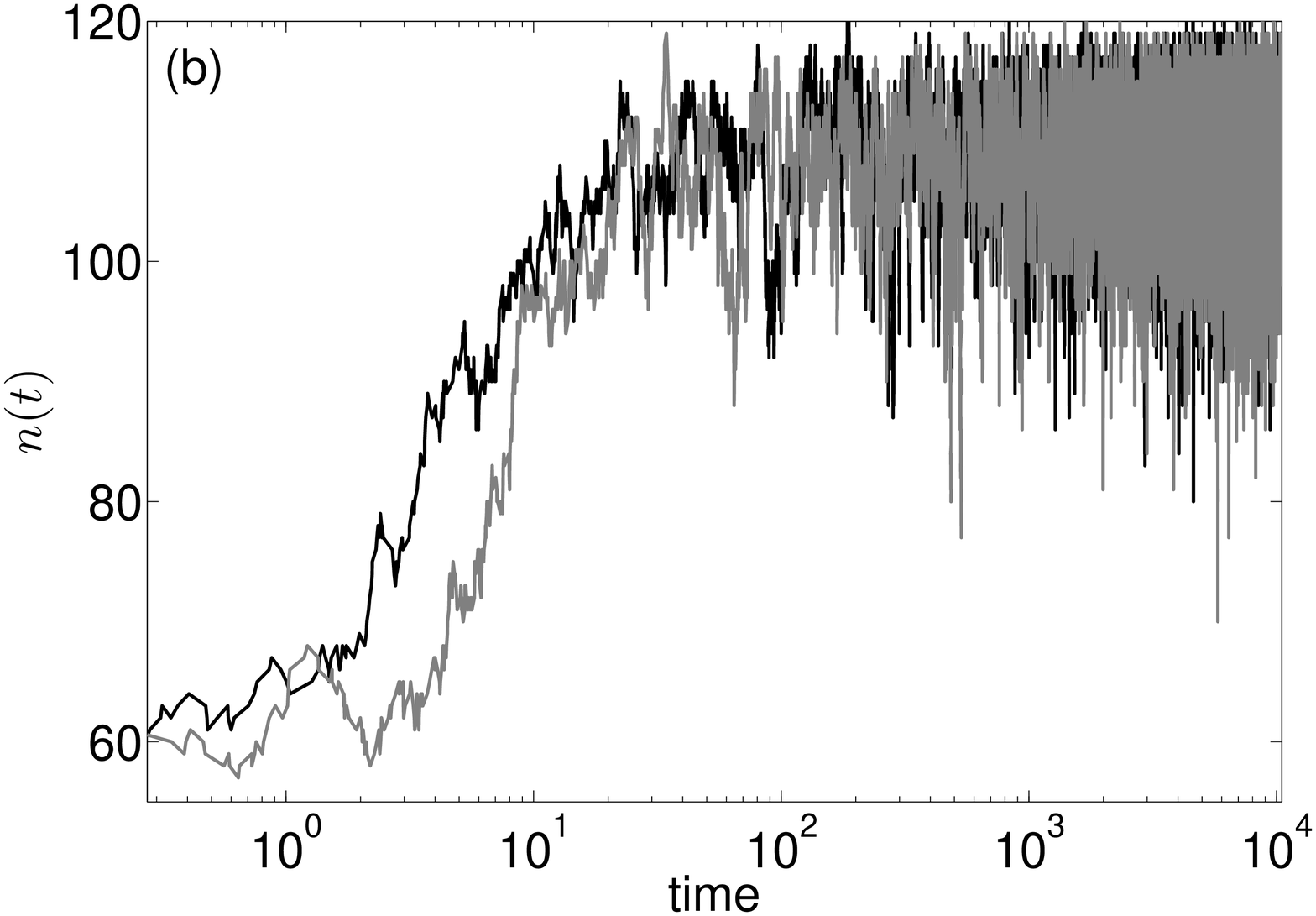}
\caption{ Number $n(t)$ of  A-susceptibles vs. time in two sample realizations (black and gray) at low zealotry, $(Z_+ + Z_-)/2N <z_c$, with initial condition $n(0)=S/2$, see text. Here, $q=2$, $N=200$ and $S=120$.
(a) Symmetric zealotry with $Z_{+}=Z_{-}=40$, $z_c=0.25$:  $n(t)$ continuously fluctuates and suddenly switches  from $n\approx S$ to $n\approx 0$ and vice-versa.  (b) Asymmetric zealotry with $Z_{+}=43$ and $Z_{-}=37$, $z_c = 0.2022$: After a transient, $n(t)$ fluctuates around a value  corresponding to a majority of  A-susceptibles, see text.}
\label{sample_fig}
\end{figure}

\begin{enumerate}
\item Pick a random voter. If this voter is a zealot nothing happens.
\item If the picked voter is a susceptible, then pick a group of $q$ neighbors (for the sake of simplicity repetition is allowed, as in Refs.~\cite{qVM,exitprobq}).
If all $q$ neighbors are in the same state, the selected voter also adopts that state.
Nothing happens in the update if there is no consensus among the $q$ neighbors~\cite{Comment}, or if the voter and its $q$ neighbors are already in the same state. 
\item Repeat the above steps {\it ad infinitum} or until consensus is reached.
\end{enumerate}
The case $q=1$ corresponds to the classical (linear) voter model~\cite{Liggett,MM1,MM2,zealot07}, and  we therefore focus on $q\geq 2$.

For the sake of  simplicity, we investigate this model on a complete graph (well-mixed population of size $N$).
The state of the population is characterized by the the probability $P(n,t)$ that the number of A-susceptibles at time $t$ is $n$. This probability obeys the master equation~\cite{noise}
\begin{eqnarray}
\label{ME}
\frac{d P_n(t)}{d t}&=&T^{+}_{n-1} P_{n-1}(t) + T^{-}_{n+1} P_{n+1}(t)
\nonumber\\
&-&  (T_n^+ + T_n^-)P_n(t).
\end{eqnarray}
The first line accounts for processes in which the number of  A-susceptibles
after the event equals $n$, while the second term accounts for
the complementary loss processes where $n\to n\pm 1$.  Here, $T_n^{\pm}$
represent the rates at which transitions occur and are given by
\begin{eqnarray}
\label{rates}
\hspace{-0.12cm}
T_n^+ = \left(\frac{S-n}{N}\right) \left(\frac{n+Z_+}{N-1}\right)^q; \;
T_n^- =  \frac{n}{N} \left(\frac{S + Z_- - n}{N-1}\right)^q
\end{eqnarray}
 When there are zealots of both types
 $T_{n=S}^+=T_{n=0}^{-}=0$  and the system has reflective boundaries at $n=0$ and $n=S$.
When there are only A-zealots, $Z_-=0$ and $Z_+=N\zeta>0$, with $\zeta$ being  the density of A-zealots, then $n=S=N(1-\zeta)$ is an absorbing boundary with
$T_{n=S}^{\pm}=0$, while $n=0$ is reflective. The birth-and-death process (\ref{ME}) is here simulated with the Gillespie algorithm \cite{Gillespie} upon rescaling time in Eq.~(\ref{ME}) as $t \to t/N$.

A quantity of particular interest is the magnetization $m=[n+Z_+ -(S-n) - Z_-]/N=
[2n-S +Z_+ -Z_-]/N$ that gives the population's average opinion or, equivalently here, the opinion of a random voter~\cite{zealot07}. We have  $m=m_{{\rm max}}= (S+Z_+-Z_-)/N$
when all susceptibles are state $+1$ (all A-susceptibles) and $m=m_{{\rm min}}=(-S+Z_+-Z_-)/N$ 
when all susceptibles are in state $-1$ (all B-susceptibles), with $m_{{\rm min}}\leq m\leq m_{{\rm max}}$.

To gain an intuitive understanding of the $q$VM dynamics, it is
useful to consider the evolution of $n(t)$ in typical sample realizations, as those  in Fig.~\ref{sample_fig} where we illustrate the dynamics at low zealotry. In Fig.~\ref{sample_fig}
we notice two distinct regimes and different time-scales. 
 In the case of symmetric low zealotry ($Z_{+}=Z_{-}$), the number of susceptibles first approaches  either the state $n\approx 0$ (all B-susceptibles)  or $n\approx S$ (all A-susceptibles). 
After a characteristic time (see Sec.~V.A), all susceptibles suddenly start switching from one state to the other, see Fig.~\ref{sample_fig} (a). A similar feature has been observed in the Sznajd model ($q=2$)
with anticonformity~\cite{genSznajd}. When $Z_{+}>Z_{-}>0$, 
the  majority of susceptibles become A-supporters  after a typical time (see Sec.~V.B). The fluctuations in the number of A-susceptibles then grow endlessly, see  Fig.~\ref{sample_fig} (b). 
An important aspect of this work is to analyze how demographic fluctuations arising in finite populations alter the mean field predictions. In Section V the  phenomena illustrated by Fig.~\ref{sample_fig} are studied in large-but-finite populations, and we show that these phenomena are beyond the reach of the next section's mean field analysis.

\section{Mean Field Description}
\begin{figure}[t]
\includegraphics[width=3.1in, height=1.5in,clip=]{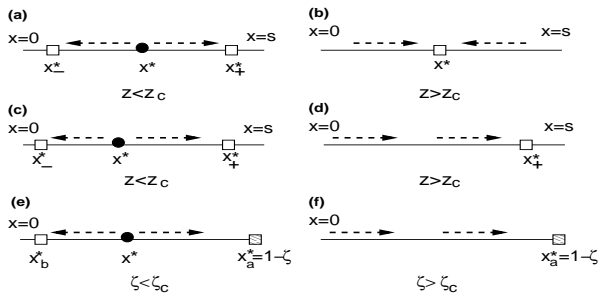}
\caption{Schematic of the mean field dynamics when $z_+=z_-=z$ and $x^*=s/2$ (a,b); when $z_{\pm}=(1\pm \delta)z>0$ with $0<\delta<1$ (c,d); and when $z_+=\zeta>0, z_-=0$ (e,f).
($\Box$) and ($\bullet$) indicate stable and unstable fixed points, respectively. 
Panels in the left column correspond to low zealotry  $z<z_c$ (a,c) and $\zeta<\zeta_{c}$ (e); those in the right column correspond to high zealotry 
$z>z_c$ (b,d) and $\zeta>\zeta_{c}$ (f), see text.}
\label{MFdiag}
\end{figure}

For further reference, it is useful to consider the mean field (MF) limit of an infinitely large population, $N\to \infty$. In such a setting,
 demographic fluctuations are negligible and the rates (\ref{rates})
can be written in terms of the density $x=n/N$ of A-susceptibles,
and the densities $z_{\pm}=Z_{\pm}/N$ of zealots of each type:
$T_n^+ \to T^+(x)= (s-x)(x+z_+)^q$ and $T_n^- \to T^-(x)= x(s+z_- -x)^q$.
The MF dynamics is  described by the rate equation
obtained by averaging $n/N$ from  Eq.~(\ref{ME})  (and rescaling  time as $Nt\to t$)~\cite{noise}:
\begin{eqnarray}
\label{eqn:MF}
\dot{x}&=& T^+(x) - T^-(x) \nonumber\\
&=&(s-x)(x+z_+)^q - x(s-x+z_-)^q,
\end{eqnarray}
where the dot denotes the time derivative and $s=S/N$.

\vspace{0.1cm}

In the absence of zealotry ($z_{\pm}=0$, $s=1$), Eq.~(\ref{eqn:MF}) has two stable absorbing fixed
points, $x=0$ (all B-supporters) and $x=1$ (all A-supporters) corresponding to 
consensus with either A or B party, separated by an unstable fixed point $x=1/2$ 
(mixture of A- and B-voters)~\cite{qVM}. It is worth noting that the dynamics of the $q$VM without zealots  ceases when a consensus is reached and this happens in 
 a finite time when the population size is finite~\cite{qVM,Sznajd,Slanina,genSznajd,vacillating}. However, in the presence of  zealots supporting both parties, the population composition endlessly fluctuates~\cite{MM2,zealot07}, see, {\it e.g.}, Fig.~\ref{sample_fig}.

\vspace{0.1cm}

In the presence of zealotry, the interior fixed points of Eq.~(\ref{eqn:MF}) satisfy   $T^+(x)=T^-(x)$, which leads to
\begin{eqnarray}
\label{eqn:FixedP}
\left(\frac{s-x}{x}\right)\left(\frac{x+z_+}{s-x +z_-}\right)^q=1.
\end{eqnarray}
Depending on the values of $z_{\pm}$ and $q$, this equation has either three physical roots,  or a single physical solution. 

\subsection{The symmetric case $z_{+}=z_{-}=z$}
When the density of zealots of both types is identical,  $z_{+}=z_{-}=z$ and $s=1-2z$ with $0<z<1/2$,
Eq.~(\ref{eqn:MF}) becomes 
\begin{equation*}
\dot{x}=(1-2z-x)(x+z)^q - x(1-z-x)^q,
\end{equation*}
that is  characterized by a fixed point $x^*=s/2$. When  $z$ is sufficiently low, 
Eq.~(\ref{eqn:MF}) has two further fixed points: $x_{+}^*$ and $x_{-}^*  = s-x_{+}^*$. 
The analysis for arbitrary $q>1$ is unwieldy, but insight can be gained by focusing on 
 $q=2$ and $q=3$, for which 
\begin{eqnarray*}
x^*_{\pm}=
\left\{
  \begin{array}{l l }
   \frac{1}{2}\left(s\pm\sqrt{1-4z}\right)  & \quad \text{($q=2$)}\\
    \frac{1}{2}\left(s\pm\sqrt{\frac{1-3z}{1+z}}\right) & \quad \text{($q=3$).}
  \end{array}\right.
\end{eqnarray*}
We readily verify that $x^*_{\pm}$  are both stable when $z<z_c(q)$, 
with $z_c(2)=1/4$ and $z_c(3)=1/3$. When $z>z_c(q)$,  the fixed points $x^*_{\pm}$ are unphysical and  $x^*=s/2$ is stable. 
This picture holds for arbitrary  finite value of $q>1$:  $x^*_{\pm}$ are stable  and the MF dynamics is characterized by bistability below a critical zealotry density  $z_c(q)$, while $x^*=s/2$ is unstable when $z<z_c$ and stable when $z\geq z_c$, see Fig.~\ref{MFdiag}(a,b). 
By determining when Eq.~(\ref{eqn:FixedP}) has three physical roots, we have found the critical zealotry density
\begin{eqnarray}
\label{eqn:zc}
z_c(q)=\left\{
  \begin{array}{l l }
   1/4  & \quad \text{($q=2$)}\\
    1/3  & \quad \text{($q=3$)}\\
3/8 & \quad \text{($q=4$)}\\
2/5 & \quad \text{($q=5$)},\\
  \end{array}\right.
\end{eqnarray}
while $z_c(1)=0$ since in the linear VM Eq.~(\ref{eqn:MF}) has always one single stable fixed point~\cite{zealot07}. Hence, the value of $z_c$  increases with $q$, while the values of $x_{+}^*$ and $x_{-}^*$ get closer to the values $0$  (all B-susceptibles) and $s$ (all A-susceptibles) as $q$ increases with $z$ kept fixed. 

In this MF picture,  the population's average opinion given by the magnetization
$m(t)=2x(t)-s$ undergoes a supercritical pitchfork bifurcation at $z=z_c$~\cite{Strogatz}: At $t\to \infty$, the critical value $z_c$ separates an ordered phase ($z<z_c$), where a majority of susceptibles supports one party, from a disordered phase ($z>z_c$) in which each party is supported by half of the susceptibles, see Fig.~\ref{MFdiag} (a,b). The stationary MF magnetization thus depends on the initial condition: when $z<z_c$,
 $m(\infty)=m^*=2x_+^* -s$ if $m(0)>0$ and $m(\infty)=-m^*$ if $m(0)<0$,
while the magnetization vanishes  when $z\geq z_c$ (or if $m(0)=0$). Using Eqs.~(\ref{eqn:FixedP}) and (\ref{eqn:zc}), it can be directly checked that just below the critical zealotry density, i.e. for $z\lesssim z_c$, the stationary magnetization is characterized by the scaling relationship $m(\infty)\propto m^*\sim \sqrt{z_c-z}$.

\subsection{The asymmetric case $z_{\pm}=(1\pm \delta)z$}
\begin{figure}
\includegraphics[width=3.5in, height=2in,clip=]{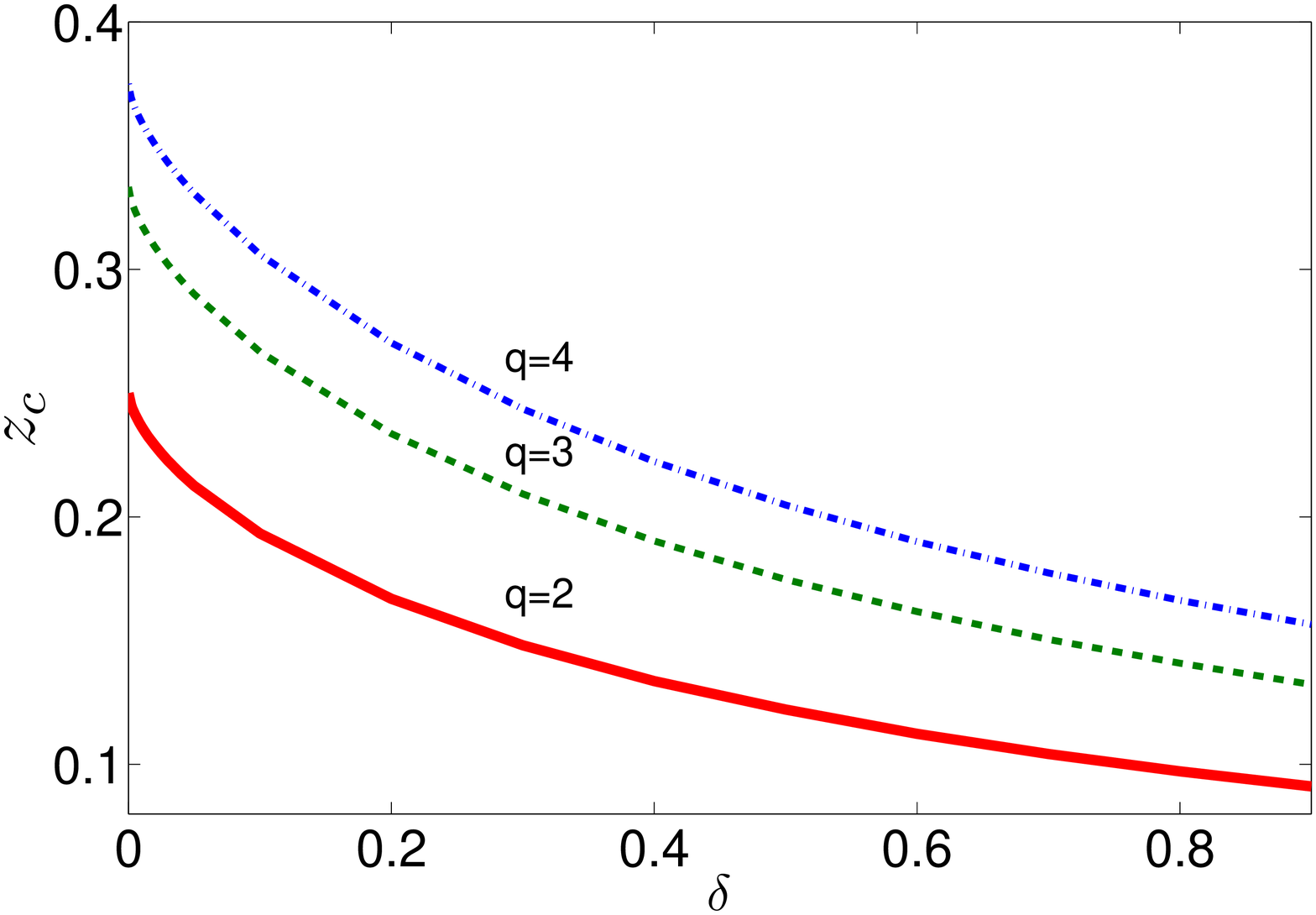}
\caption{(\textit{Color Online}) Critical value of $z_c$ as a  function $\delta=(z_+ - z_-)/2z$ for $q=2$ (solid),
$q=3$ (dashed), and  $q=4$ (dash-dotted) in the case of asymmetric zealotry.
There is bistability  where $z=(z_+ - z_-)/(2\delta)<z_c$,
see text.}
\label{ZCvsDelta}
\end{figure}

When the number of A-zealots  exceeds that of B-zealots, with 
$z_{+}>z_{-}>0$, it is convenient to use the parametrization 
\begin{equation}
\label{param}
z_{\pm}=(1\pm \delta)z, 
\end{equation}
where $\delta=(z_+ - z_-)/2z$ quantifies the zealotry asymmetry.
With Eq.~(\ref{param}), we still have $s=1-2z$ with $0<z<1/2$ and  Eq.~(\ref{eqn:MF}) becomes
\begin{equation*}
\dot{x}=(1-2z-x)[x+(1+\delta)z]^q - x[1-(1+\delta)z-x]^q.
\end{equation*}
This rate equation is  also characterized by bistability at low zealotry, with two stable fixed points  $x_{\pm}^*$ separated by an unstable fixed point $x^*$, and by  the sole stable fixed point $x_+^*$ at higher zealotry, see Fig.~\ref{MFdiag}(c,d).
By determining when  Eq.~(\ref{eqn:MF}) has three physical fixed points, we have determined the critical density of zealotry $z_{c}(q,\delta)$, see Fig.~\ref{ZCvsDelta}:  At fixed $q$ and $\delta$, the fixed points $x_{\pm}^*$ are stable when $z<z_c$ while only $x_+^*$ is stable when $z\geq z_c$. We have found that $z_{c}$ decreases with $\delta$ (at fixed $q$) and increases with $q$ (at fixed $\delta$). 

In this MF picture, the opinion of a random individual  is given by the magnetization $m(t)=2(x+\delta z)-s$. The critical zealotry density $z_c$
separates a bistable phase ($z<z_c$) from a phase where most susceptibles support the party having more zealots, see Fig.~\ref{MFdiag} (c,d). Hence, the stationary MF magnetization at low zealotry ($z<z_c$)
depends on the initial condition and is $m(\infty)=m_+^*=2(x_+^* + \delta z)-s$ if $m(0)>2(x^* + \delta z)-s$ and $m(\infty)=m_-^*=2(x_-^* + \delta z)-s$ if $m(0)< 2(x^* + \delta z)-s$. When $z>z_c$ the 
stationary MF magnetization is  $m(\infty)=m_+^*$.

\subsection{The absorbing case $z_{+}=\zeta, z_{-}=0$}
When there are only A-zealots, $z_+=\zeta>0$ and $z_-=0$,
 Eq.~(\ref{eqn:MF}) becomes
\begin{equation*}
\dot{x}=(1-\zeta-x)\left[(x+\zeta)^q - x(1-\zeta-x)^{q-1}\right],
\end{equation*}
and has an absorbing fixed point  $x^*_a=1-\zeta$.
Below a  critical zealotry density $\zeta_c(q)$, this rate equation admits  two other fixed points: $x^*_b$, that is stable, and $x^*$ that is unstable and separates $x^*_a$ and $x^*_b$, see Fig.~\ref{MFdiag}(e,f). When $\zeta>\zeta_c(q)$, the absorbing state $x^*_a=1-\zeta$ is the only fixed point. For $q=2$ and $q=3$, we explicitly find
\begin{eqnarray}
\label{xb}
x^*_{b}=
\left\{
  \begin{array}{l l }
   \frac{1}{4}\left(1-3\zeta-\sqrt{1-(6-\zeta)\zeta}\right)  & \quad \text{($q=2$)}\\
    \frac{1-2\zeta(1+\zeta)-\sqrt{1-4\zeta}}{2(2+\zeta)} & \quad \text{($q=3$)}
  \end{array}\right.
\end{eqnarray}
and 
\begin{eqnarray}
\label{xu}
x^*=
\left\{
  \begin{array}{l l }
   \frac{1}{4}\left(1-3\zeta+\sqrt{1-(6-\zeta)\zeta}\right)  & \quad \text{($q=2$)}\\
    \frac{1-2\zeta(1+\zeta)+\sqrt{1-4\zeta}}{2(2+\zeta)} & \quad \text{($q=3$)}
  \end{array}\right.
\end{eqnarray}
From these expressions, and more generally by determining when Eq.~(\ref{eqn:MF}) has three physical fixed points, we have found the critical zealotry density in the absorbing case:
\begin{eqnarray}
\label{eqn:zcp}
\zeta_{c}(q)=\left\{
  \begin{array}{l l }
   3-2\sqrt{2}  & \quad \text{($q=2$)}\\
    1/4  & \quad \text{($q=3$)}\\
0.295 & \quad \text{($q=4$)}\\
0.326 & \quad \text{($q=5$)}\\
  \end{array}\right.
\end{eqnarray}
We thus distinguish two regimes:

(i) When $\zeta< \zeta_{c}(q)$ both $x_{a,b}^*$ are stable and the dynamics crucially depends on the initial density $x_0$ of A-susceptibles: 
If $x_0>x^*$, the final state is the consensus with party A; whereas the steady state consists of a vast majority of B-party voters  when  
$x_0<x^*$. In Sec. VI, we show that random fluctuations drastically alter this picture: In a finite population, $x_{b}^*$ is a metastable state when $\zeta<\zeta_{c}(q)$ and $x_0<x^*$, and we shall see that  the A-consensus is  reached after a very long transient that scales exponentially with the population size.

(ii) When $\zeta > \zeta_{c}(q)$, as well as when $\zeta=\zeta_c$ and $x_0>x^*$, the absorbing state is rapidly reached. 

\section{Stationary probability distribution}
In this section, we compute the stationary probability distribution (SPD) of the $q$VM with zealotry when there is no absorbing state, and show that it shape generally differs from the Gaussian-like distribution obtained in the linear VM with zealots~\cite{zealot07}. 

\vspace{0.1cm}

The SPD $P_n^*=\lim_{t\to \infty}P_n(t)$ obeys the following stationary  master equation, obtained from Eq.~(\ref{ME}):
\begin{equation*}
T^{+}_{n-1} P_{n-1}^* + T^{-}_{n+1}P_{n+1}^* -(T_n^+ + T_n^-)P_{n}^*=0.
\end{equation*}
The exact SPD is uniquely  obtained by iterating  the detailed balance relation  $T_{n-1}^+ P_{n-1}^*=T_n^- P_{n}^*$~\cite{noise}, yielding
\begin{eqnarray}
\label{PnS}
\hspace{-5mm}
P_n^*&=&P_0^*~\prod_{j=0}^{n-1} (T_j^+/T_{j+1}^-)\nonumber\\&=&
P_0^*~\prod_{j=0}^{n-1} \left(\frac{S-j}{j+1}\right)\left(\frac{j+ Z_{+}}{S+Z_{-}-j-1}\right)^q,
\end{eqnarray}
where the normalization $\sum_{n=0}^{S}P_n^*=1$ gives
$P_0^*=1/[1+\sum_{k=1}^{S} \prod_{j=0}^{k-1} (T_j^+/T_{j+1}^-)]$ and $P_S^*=1-P_0^*-\sum_{k=1}^{S-1} P_n^*$.

Since   $n=N[(m+s)/2 -\delta z]$, the stationary magnetization  distribution  $Q_m^*$ 
has the same shape as $P_n^*$, with 
\begin{eqnarray}
\label{QmS}
Q_m^*&=&P_{N[(m+s)/2 -\delta z]}^*\\&=&P_0^*~\prod_{j=0}^{N[(m+s)/2 -\delta z]-1} \left(\frac{S-j}{j+1}\right)\left(\frac{j+ Z_{+}}{S+Z_{-}-j-1}\right)^q.\nonumber
\end{eqnarray}
%

In large populations, a useful approximation of (\ref{PnS}) is obtained by writing $P_n^*=
P_0^*~{\rm exp}\left(\sum_{j=0}^{n-1} \Psi_j \right)$
with $\Psi_j=\ln{(T_j^+/T_{j+1}^-)}$, and by using Euler-MacLaurin formula 
$\sum_{j=0}^{n-1} \Psi_j = \int_{0}^{n-1} \Psi_j~dj + (\Psi_0 + \Psi_{n-1})/2$, 
where we have neglected higher order terms~\cite{Arfken}.

When $N\gg 1$,
it is  useful to work in the continuum limit with the rates $T_n^{\pm}\to T^{\pm}(x)$, as in Sec.~III. 
By introducing 
\begin{eqnarray}
\label{Psi}
\Psi(x)=\ln{[T^+(x)/T^-(x)]},
\end{eqnarray}
we have $ \sum_{j=0}^{n-1} \Psi_j \simeq N\int_{0}^{x} \Psi(x)~dx$ to leading order in $N$.
  Hence, the leading contribution to the SPD when $N\gg 1$ is 
\begin{eqnarray}
\label{eqn:PS}
P_{n}^* 
&\sim& P_0^*~ {\rm exp}\left(N~\int_{0}^{x}\Psi(y)~dy\right)= P^*(x). 
\end{eqnarray}

The local extrema of $P^*(x)$ satisfy $\Psi(x)=0$, see (\ref{Psi}), and thus coincide with the fixed points of Eq.~(\ref{eqn:MF}). As a consequence, in large populations $P_n^*$ is either characterized by a single peak at $n^*=Nx^*$ when $z> z_c$, or has two  peaks at the metastable states $n_{\pm}^*= Nx_{\pm}^*$ when $z<z_c$. In this case, there is bistability and the amplitudes of the peaks at $n_{\pm}^*$ are in the ratio ($N\gg 1$)
\begin{eqnarray}
\label{peaks}
\frac{P_{n_{+}^{*}}^*}{P_{n_{-}^{*}}^*}\sim e^{N\int_{x_{-}^*}^{x_{+}^*}\Psi(y)~dy}.
\end{eqnarray}
The integrals in Eqs.~(\ref{eqn:PS}) and (\ref{peaks}) can be computed, but their expressions are  unenlightening. Here, we infer  the properties of $P_n^* \sim P^*(x)$ and $Q_m^*$ from those of $\Psi(x)$.
\subsection{Stationary probability distribution in the symmetric case}
In the symmetric case, $z_+=z_-=z$, Eq.~(\ref{Psi}) becomes
\begin{equation*}
\Psi(x)=\ln{\left[\left(\frac{1-(x+2z)}{x}\right)\left(\frac{x+z}{1-(x+z)}\right)^q\right]}
\end{equation*}
and has the symmetry $\Psi(x)=-\Psi(s-x)$.
We distinguish the cases of low and high zealotry density:

\vspace{0.15cm}

\begin{figure}
\includegraphics[width=3.6in, height=2.0in,clip=]{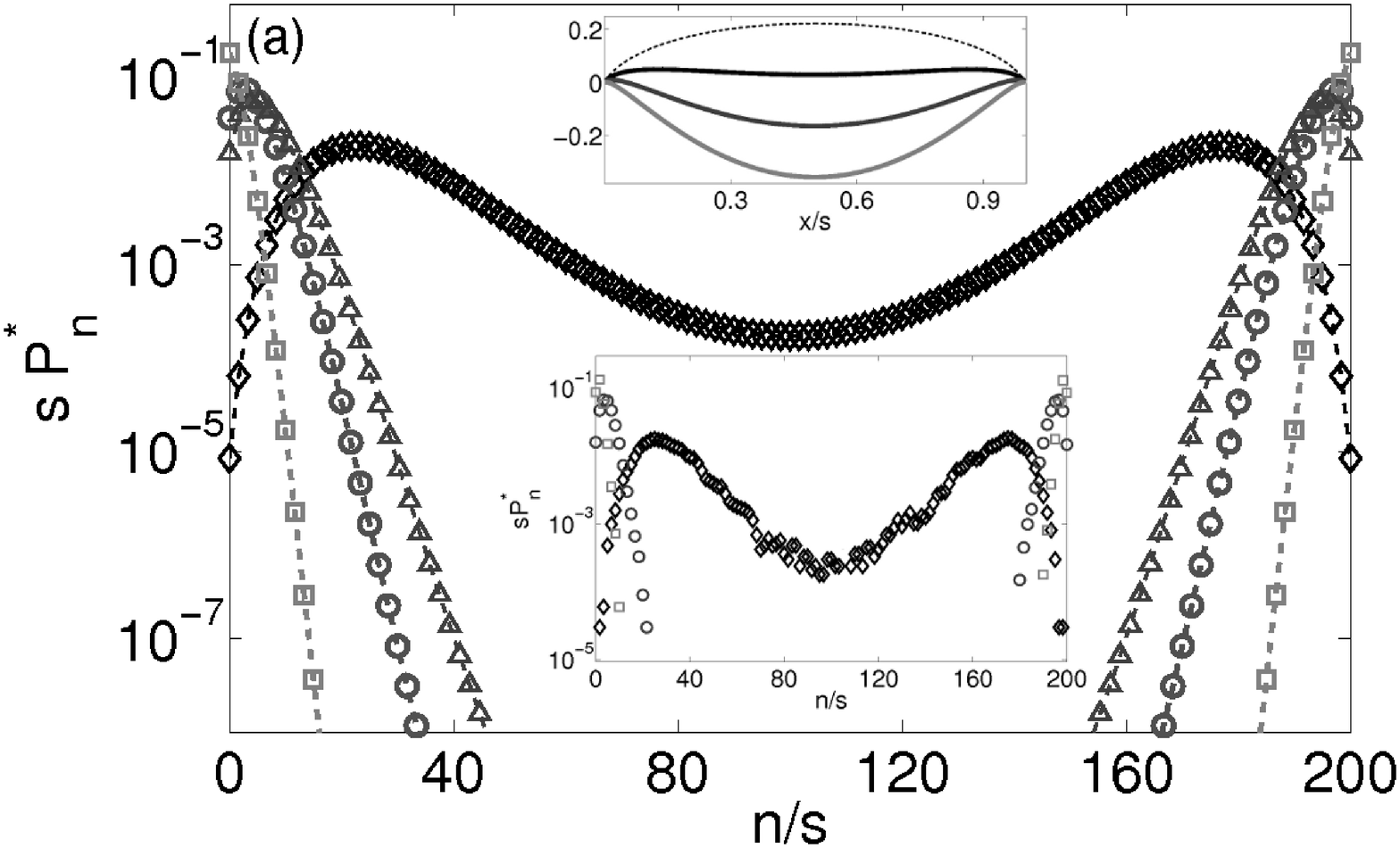}
\includegraphics[width=3.6in, height=2.0in,clip=]{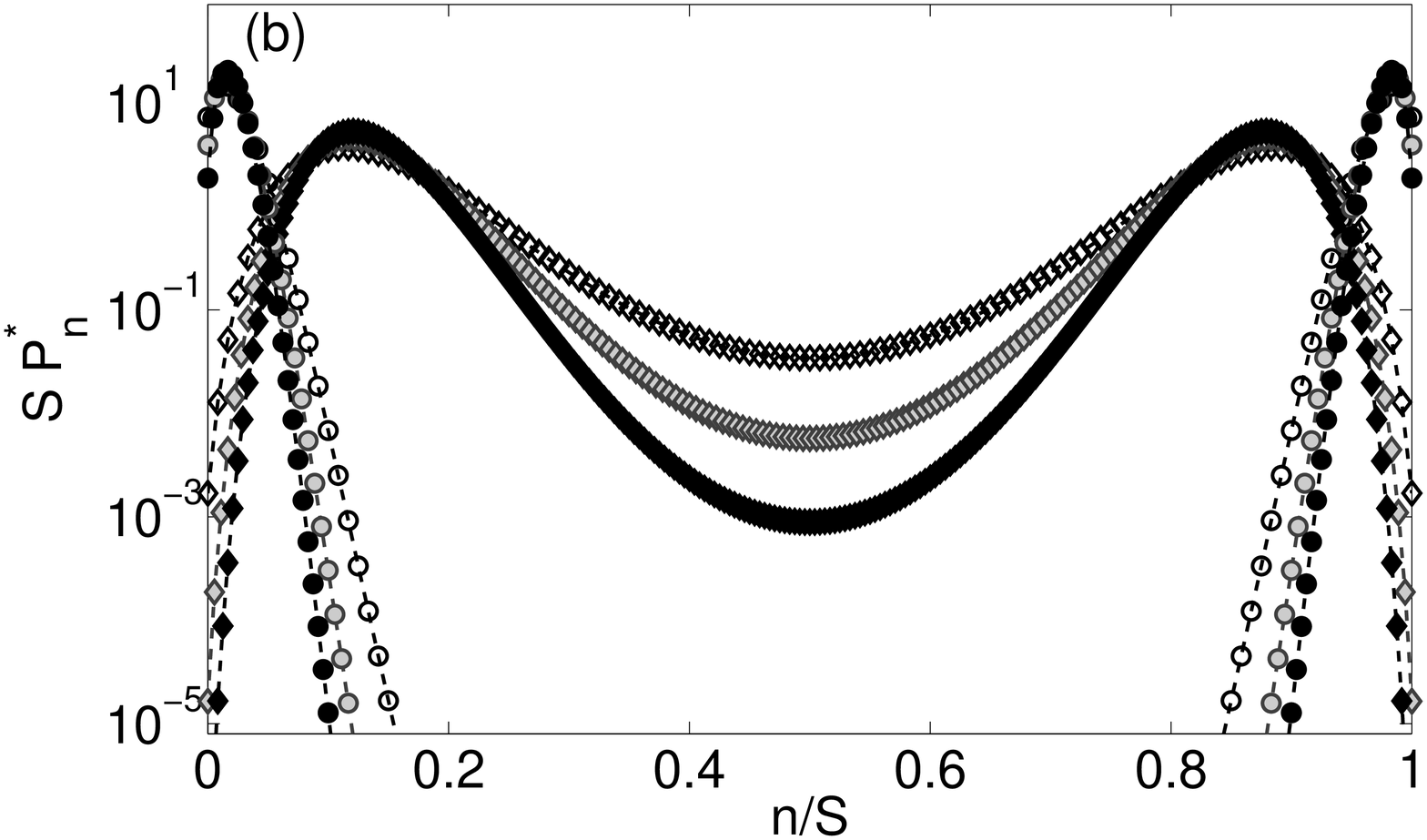}
\caption{ 
Rescaled SPD at low zealotry  with $z_+=z_-=z<z_c$ in semi-log scale.
(a) $sP_n^*$ vs. $n/s$ from Eq.~(\ref{PnS}) for different values of $q$ and $z$, with $N=200$. Here,  
$q=2,z=0.2$ ($\Diamond$); $q=3, z=0.22$ ($\Delta$); $q=3,z=0.2$ ($\circ$)  and $q=4,z=0.2$ ($\Box$). 
Lower inset: Similar; $sP_n^*$ vs. $n/s$ from stochastic simulations. Upper inset: $\int_0^x~\Psi(y)~dy$ vs. $x/s$ for $z=0.2$ and, from top to bottom, $q=1$  (dashed), $q=2$ (black), $q=3$ (gray) and $q=4$ (light gray).
 (b) $SP_n^*$ vs. $n/S$ from Eq.~(\ref{PnS}) for $q=2$ (diamonds) and $q=3$ (circles), and for different values of the population size $N$ at low zealotry. Here,  $z=0.2$ and $N=200$ (open symbols), $N=300$ (gray-filled symbols) and $N=400$ (black-filled symbols). Not shown in panels (a) and (b)
is the  range where  $P_n^*\lesssim 10^{-8}$ (where $P_n^*\lesssim 10^{-5}$ in the lower inset).
}
\label{Fig4}
\end{figure}

(i) When $z<z_c(q)$, the fixed points $x^*$ and $x_{\pm}^*$ of Eq.~(\ref{eqn:MF}) 
are also the roots of $\Psi(x)$. 
Hence, when $N\gg 1$,  $P_{n}^*=P_{S-n}^*\sim P^*(x) \propto e^{N\int_0^x \Psi(y) dy}$ is a symmetric bimodal SPD characterized by two peaks at $n= n_{\pm}^*$. As a consequence,  $Q_m^*=Q_{-m}^*$ is an even function.

\begin{figure}
\includegraphics[width=3.6in, height=2.0in,clip=]{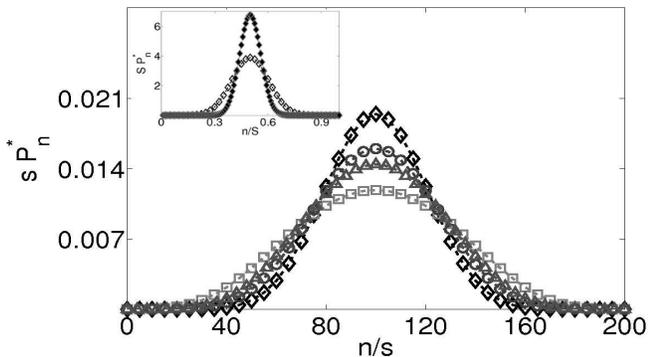}
\caption{
 $sP_n^*$ vs. $n/s$ for $q=2 - 4$ at high zealotry $z>z_c$.
 Here,   $N=200$ and $(q,z)=(2,0.4)$ ($\Diamond$), $(3,0.4)$ ($\circ$), $(3,0.375)$ ($\Delta$), $(4,0.4)$ ($\Box$).
The SPDs  have a single peak  at $n/s=N/2$ and width broadens when $q$ and $1/z$ are increased at fixed $N$. 
Inset: $SP_n^*$ vs. $n/S$
for $q=2, z=0.4$ and different values of $N$. Here, $N=200$ ($\Diamond$) and $N=600$ ($\blacklozenge$).
}
\label{Fig5}
\end{figure}

In Figure~\ref{Fig4} (a), we show the exact  SPD for $q=2-4$ 
characterized by two peaks of same intensity  at $n=n_{\pm}^*$ and a local minimum at $n^*=S/2$.  We remark that when $q$ is increased, the SPD vanishes dramatically away from the peaks. In fact, since $\int_0^x \Psi(y) dy$  is close to zero or negative on $ x_{-}^* \ll x\ll x_{+}^*$,
see Fig.~\ref{Fig4} (a, upper inset), $P_{n_-^*\ll n\ll n_+^*}^*$ vanishes exponentially with $N$ and when $q$ is increased. 
Fig.~\ref{Fig4}(a) shows that the SPD steepens and its peaks are more pronounced when $q$ and $1/z$ are increased and $N$ is kept fixed. We have also obtained the (quasi-)SPD from stochastic simulations, see 
Fig.~\ref{Fig4} (a, lower inset),  by  averaging over $25,000$ realizations after $40,000$ simulation 
steps. While unavoidably more noisy, the simulation results reproduce the predictions of Eq.~(\ref{PnS}). Fig.~\ref{Fig4}(b)  shows how  $SP_{n}^*$ scales with
$n/S=x/s$ for different population sizes, and we notice that the main influence of raising $N$ is to concentrate the probability density $SP_{n}^*$ around the peaks whose location are essentially unaffected by $N$ (when $N\gg 1$).
In Fig.~\ref{Fig4}, we also notice that the symmetric peaks are clearly identifiable when $q=2$ and $q=3$, but almost coincide with $n=0$ and $n=S$ for $q=4$. This is because $x_{\pm}^*$ approach the values
$x=0,s$ when $q$ is increased. 

(ii) When $z\geq z_c$, the only physical root of $\Psi(x)$  is  $x^*=s/2$, as in the  classical voter model~\cite{zealot07}. Hence,  
$P_{n}^*= P_{S-n}^*\sim P^*(x) \propto e^{N\int_0^x \Psi(y) dy}$ 
has a single maximum at $x=s/2$ when $N\gg 1$.
The resulting symmetric Gaussian-like distribution centered at 
$n^*=S/2$ when $N\gg 1$, see Fig.~\ref{Fig5}, is very similar to the SPD obtained in the classical voter model with zealots~\cite{zealot07}. Fig.~\ref{Fig5}(inset) illustrates that the probability density
steepens around $s/2$ when the population size is increased.

\subsection{Stationary probability distribution in the asymmetric case}
In the asymmetric case, with zealot densities $z_{\pm}=(1\pm \delta)z$ and $\delta>0$,  Eq.~(\ref{Psi}) is
\begin{equation*}
\Psi(x)=\ln{\left[\left(\frac{1-(x+2z)}{x}\right)\left(\frac{x+z(1+\delta)}{1-z(1+\delta)-x}\right)^q\right]}
\end{equation*}
and has either three or one physical roots:

\vspace{0.15cm}

\begin{figure}
\includegraphics[width=3.6in, height=2.0in,clip=]{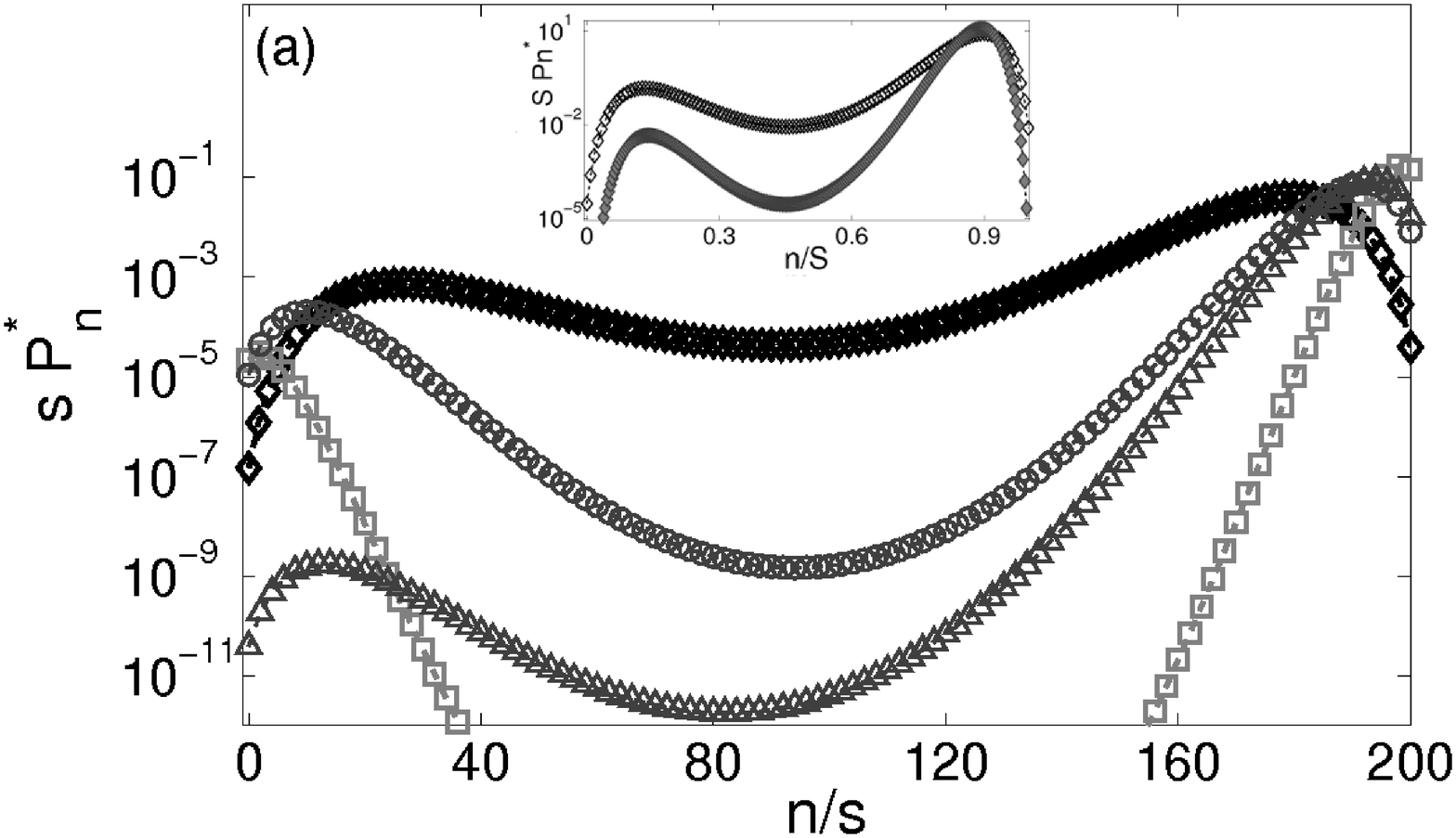}
\includegraphics[width=3.6in, height=2.1in,clip=]{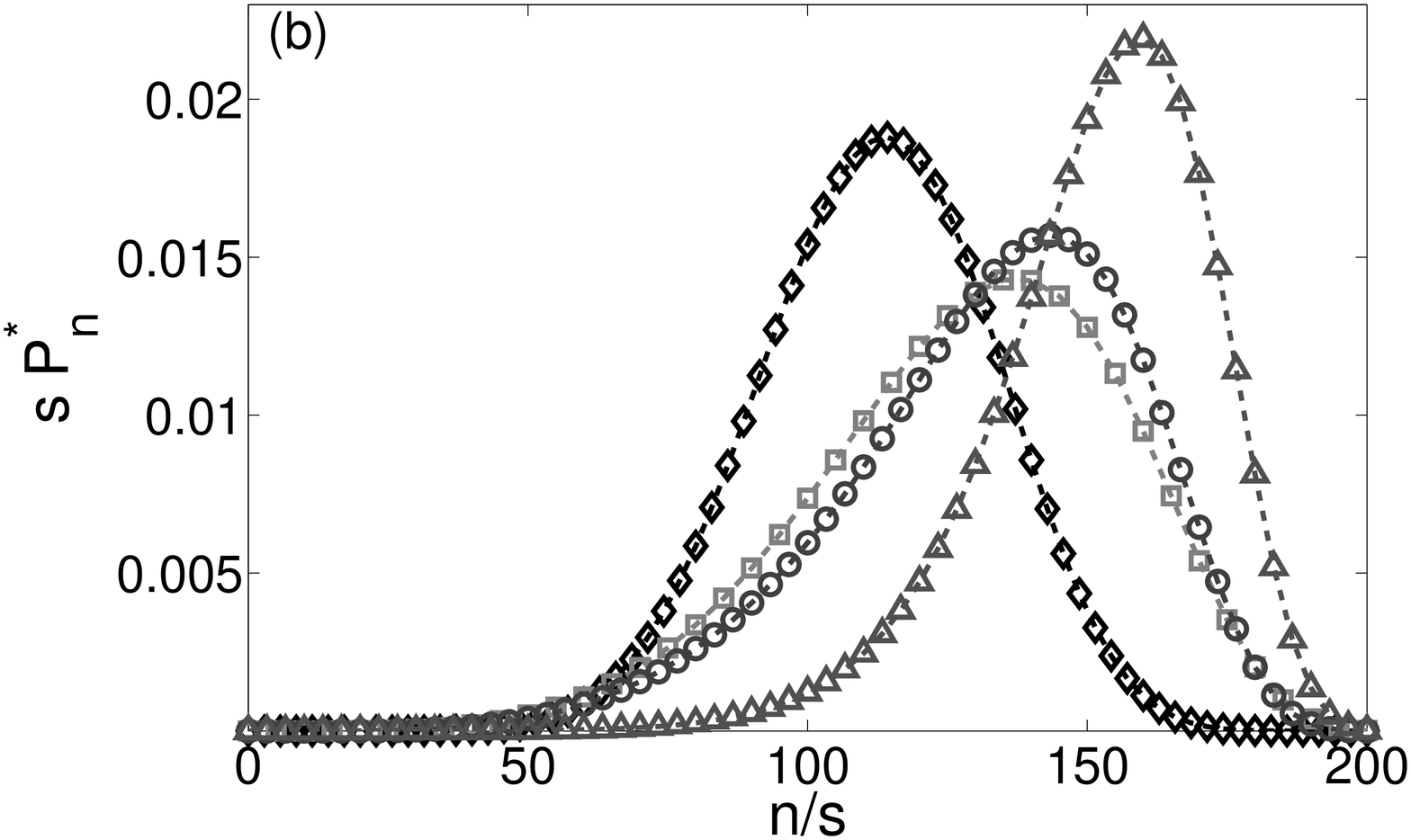}
\caption{
Rescaled SPD under asymmetric zealotry 
$Z_{\pm}=N(1\pm \delta)z$.
(a)
 $sP_n^*$ vs. $n/s$ from Eq.~(\ref{PnS}) at low zealotry ($z<z_c$). Here, $N=200$ and 
$(q,Z_+,Z_-)=(2,41,39)$ ($\Diamond$), $(3,51,49)$ ($\circ$), $(3,53,47)$ ($\Delta$), $(4,51,49)$ 
($\Box$).%
 For $q=4$, the range where  $P_n^*\lesssim 10^{-12}$ is not shown. Inset: $SP_n^*$ vs. $n/S$ for $q=2$ and $(N,Z_+,Z_-)=(200,41,39)$ (open $\Diamond$), $(400,82,78)$ (gray-filled symbols).
 (b) Left-skewed rescaled SPD at high zealotry,  with a single peak 
at $n_{+}^*$.
Here,  $N=200$ and
$(q,Z_+,Z_-)=(2,67,63)$ ($\Diamond$), $(3,72,68)$ ($\circ$), $(3,74,66)$ ($\Delta$), $(4,82,78)$ 
($\Box$), see text.
}
\label{Fig6}
\end{figure}

(i) At fixed $q$ and $\delta$, when $z<z_c(q,\delta)$, the fixed points $x^*$ and $x_{\pm}^*$ of Eq.~(\ref{eqn:MF}) 
are the physical roots of $\Psi(x)$. 
Since $P_{n}^* \sim P^*(x) \propto e^{N\int_0^x \Psi(y) dy}$ when $N\gg 1$, 
the SPD is again a bimodal distribution peaked at $n_{\pm}^*$. However, $x_{+}^*$ has a greater basin of attraction than $x_{-}^*$ and $\int_{x_{-}^*}^{x_{+}^*} \Psi(y)~dy>0$. As a consequence, the SPD is asymmetric, with the peak  at $n_{+}^*$ being much stronger than the one at $n_{-}^*$. The ratio of the peaks is given  by (\ref{peaks}), which shows that the asymmetry of $P_{n}^*$  grows exponentially with $N$ and increases with $q$, see Fig.~\ref{Fig6}(a). 
While  an asymmetry in the zealotry in the linear VM  does not significantly affect the form of the SPD~\cite{zealot07}, we here find that in  the $q$VM even a small bias in the  zealotry drastically changes the shape of the SPD and leads to marked dominance of the party with more zealots.

In Fig.~\ref{Fig6} (a), we report the exact  SPD for $q=2-4$ and illustrate its asymmetric bimodal nature, with marked peaks of different intensities at $n_{\pm}^*$. We notice that the 
asymmetry in the peaks intensity, given by (\ref{peaks}), is stronger when we increase $q$ and $Z_+ - Z_-\propto \delta z$. As in the symmetric case, the SPD decays dramatically away from the peaks and  $P_{n_-^*\ll n\ll n_+^*}^*$ vanishes with $N\gg 1$ and when $q$ is increased. In Fig.~\ref{Fig6} (a, inset)
we show that  the SPD remains bimodal  when the population size is increased, and the main influence of raising $N$ is to concentrate the probability density near its peak at  $x_{+}^*=n_{+}^*/N$ (when $N\gg 1$).

(ii) At fixed $q$ and $\delta$, when $z>z_c$, the only real root of $\Psi(x)$  is  $x=x_{+}^*$. This lies closer to $x=s$ than to $x=0$, and hence $\int_0^x \Psi(y) dy$  
is an asymmetric function with a  single maximum at $x=x_{+}^*$. Therefore, in large populations $P_{n}^*$ is an asymmetric left-skewed SPD with a single peak at $n=n_+^*$, as shown in Fig.~\ref{Fig5}(b)
where we see that the SPD broadens when $q$ is increased and that it steepens
when $Z_+ -Z_-\propto \delta z$ is increased. 
As above, the probability density
steepens around  $x_{+}^*$ when $N$ is increased.

\section{Fluctuation-driven dynamics at low zealotry}
We now study how a small non-zero density of  zealots of both parties ($0<z<z_c$)
affects the $q$VM long-time dynamics. We show that, after a typically long transient, all susceptibles voters switch allegiance from 
the state $n=0$ (all B-susceptibles) to  state $n=S$ (all A-susceptibles) in a typical switching time.
In the symmetric case, there is ``swing-state dynamics'' with all susceptibles endlessly swinging allegiance. In the asymmetric case where party A has more zealots than party B, the dynamics is characterized by various time-scales and by growing fluctuations around the metastable state $n_+^*$. 
Below,  we  show that 
the long-time $q$VM dynamics is  driven by fluctuations and characterized by a mean switching time that scales (approximately) exponentially with the system size $N$ in large-but-finite populations.

\subsection{Swing-state dynamics and switching time in the case of symmetric zealotry}
\begin{figure}
\begin{center}
\includegraphics[width=3.6in, height=2in,clip=]{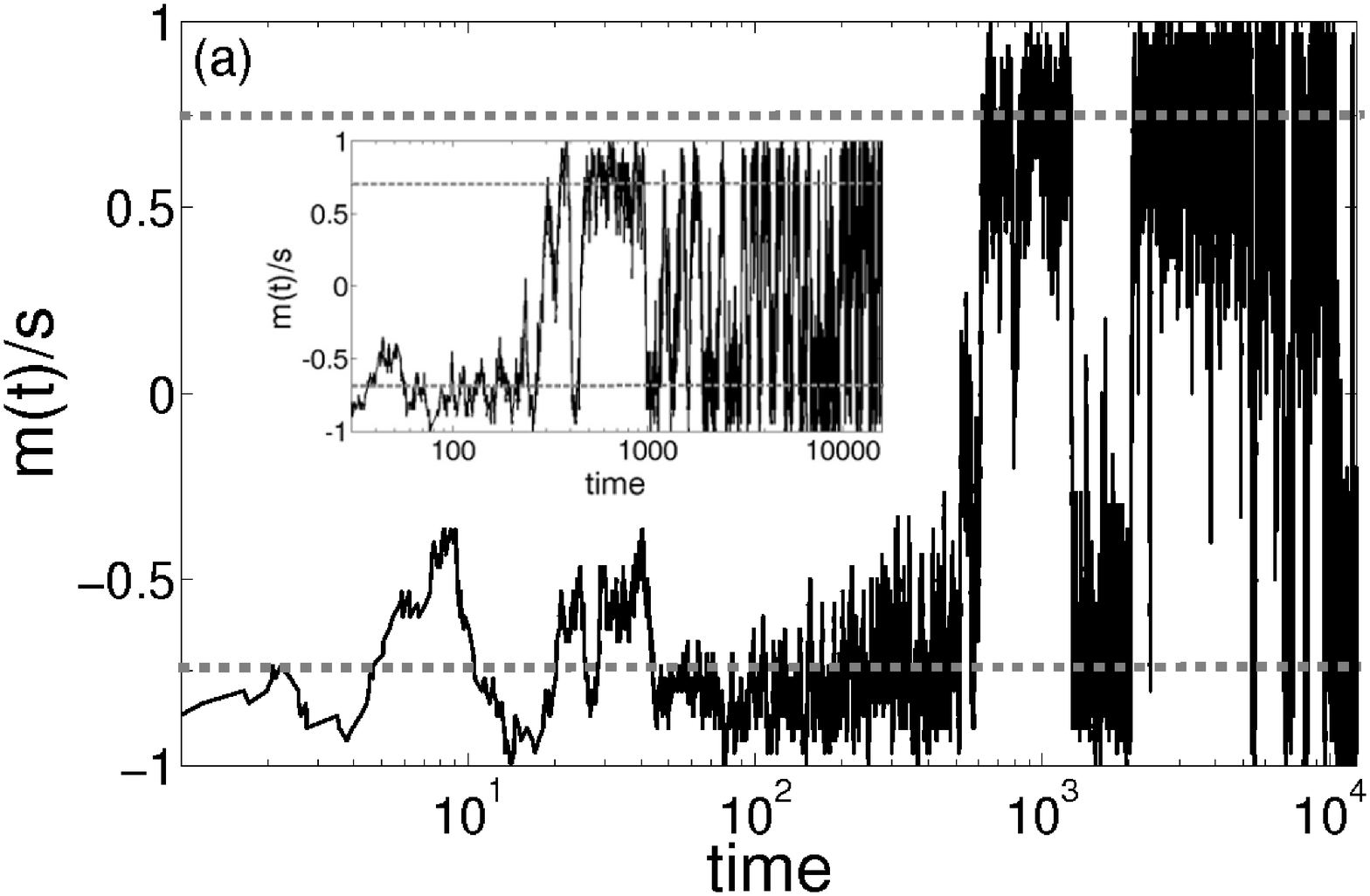}
\includegraphics[width=3.7in, height=2.2in,clip=]{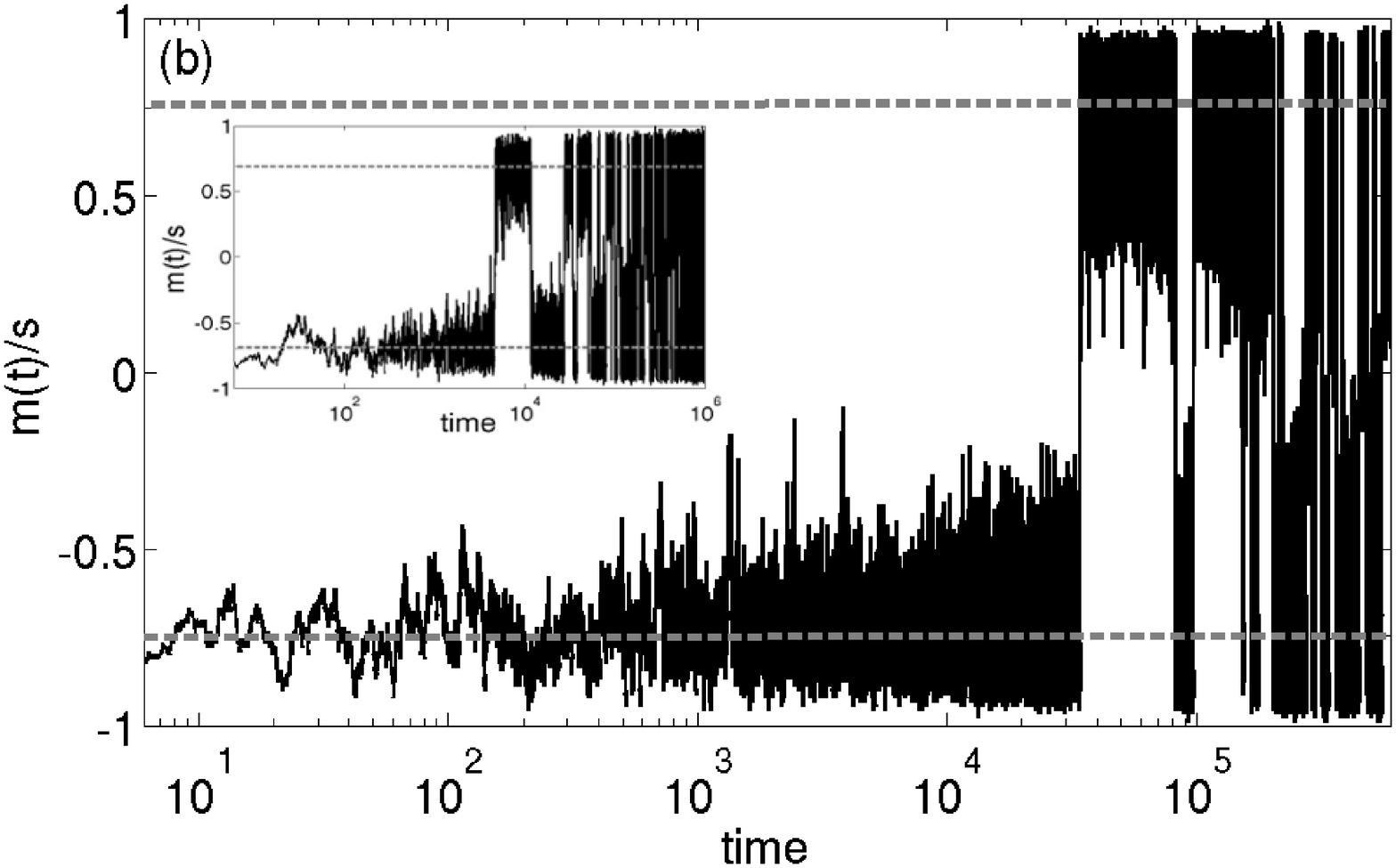}
\caption{
Typical evolution of the  rescaled magnetization for different values
of $q$ and $N$ with initial condition $m(0)=-s$ (all B-susceptibles) on a semi-log scale:
(a) Single realization of  $m(t)/s$   for  $q=2, z=z_+=z_-=0.2<z_c$ and population size $N=100$.
 At time $t\approx 625$,  $m=s$ and starts swinging back and forth the values  $m=\pm s$. Here, the MF predicts  $m^*/s\approx 0.745$ and $\pm m^*/s$  are shown as dashed lines. Inset: $m(t)/s$ vs. time for  $q=3, z=0.3<z_c$ and $N=100$. The system starts swinging between $m=\pm s$
at $t\approx 640$. Here, $\pm m^*/s\approx \pm 0.693$ (dashed).
(b) $m(t)/s$ vs. time for the same parameters as in (a) but with $N=300$. The system's magnetization switches to $m=s$ only at $t\approx 2\cdot 10^5$.
Inset: $m(t)/s$ vs. time for  $q=3, z=0.3$ and $N=400$. The magnetization switches to $m=s$ at $t\approx 3\cdot 10^5$. The difference in the switching times in (a) and (b) results from the exponential scaling  of the 
mean switching time on $N$, see text.
}
\label{Fig7}
\end{center}
\end{figure}
As illustrated in Fig.~\ref{sample_fig}(a), the long-time dynamics in the symmetric case
 is characterized by the continuous swinging from states $n\approx 0$ to 
$n\approx S$ and vice versa. When $Z_+=Z_-$, all susceptibles thus continuously switch allegiance in the long run. In that regime,  the magnetization $m(t)=(2n(t)-S)/N$ is thus characterized by  abrupt jumps from $m\approx \pm s$ to $m\approx \mp s$, see Fig.~\ref{Fig7}, 
while the stationary ensemble-averaged magnetization
 $\langle m(\infty) \rangle =\sum_{m=-s}^{s} m Q_m^*=0$, since 
$Q_m^*$ is even and 
each agent is as likely to be in one or the opposite state. A similar phenomenon has been found in the Sznajd model ($q=2$)
 with anticonformity~\cite{genSznajd}.

  This swing-state phenomenon is not captured by the mean field description of Sec.~III and is here 
characterized by   the mean time  $\tau_{0}^{S}$ to switch for the first time from state $n=0$  to
 $n=S$. The scaling of $\tau_{0}^{S}$ on $N$ allows us to rationalize the data of Fig.~\ref{Fig7} where
the switching time is found to dramatically increase with the population size.  Clearly, the symmetry implies that the mean switching, or swinging, time $\tau_{0}^{S}$ is identical to the mean time  $\tau_{S}^{0}$ to switch from   $n=S$ to $n=0$.

Finding the mean switching time  can  be formulated as a first-passage time problem and, when $N\gg 1$, $\tau_0^S$ can  be computed  using the framework of the backward Fokker-Planck equation (bFPE)~\cite{noise}. In this context, the model's bFPE infinitesimal generator is 
\begin{eqnarray}
\label{G}
\hspace{-2mm}
{\cal G}_{{\rm b}}(x)=[T^+(x)-T^{-}(x)]\partial_x +
\frac{[T^+(x)+T^{-}(x)]}{2N}\partial_x^2.
\end{eqnarray}
The mean time $\tau^S(x_0)$ to be absorbed at $x=s$ (all A-susceptibles), starting  from the initial state 
$x=x_0$, with a  reflective boundary at $x=0$ (all B-susceptibles),  obeys
\begin{eqnarray}
\label{bFPE}
{\cal G}_{{\rm b}}(x_0)~\tau^S(x_0)=-1,
\end{eqnarray}
with $(d/dx)\tau^S(0)=0$  and $\tau^S(s)=0$ (reflective and absorbing boundaries)~\cite{noise,KramersReview}.
To obtain the mean switching time $\tau_0^S$ we solve  Eq.~(\ref{bFPE}) with $x_0=0$ using standard methods~\cite{noise},
and obtain
\begin{eqnarray}
\label{TOS1}
\tau_0^S= 2N\int_0^{s} dy~e^{-N\phi(y)}~\int_0^{y} \frac{e^{N\phi(v)}~dv}{T^+(v) + T^- (v)},
\end{eqnarray}
where $\phi(v)=-2\int_0^v du~\left\{\frac{T^-(u) - T^+ (u)}{T^-(u) + T^+ (u)}\right\}$.
As with other fluctuation-driven phenomena 
associated with metastable states, see {\it e.g.} \cite{Kramers,KramersReview,3SVMZ,WKB,otherWKB} and below, this result predicts that the mean switching time $\tau_0^S$ grows (approximately) exponentially with the population size $N$. This explains the difference of various orders of magnitude in the switching time observed in  Figs.~\ref{Fig7}(a) and 7(b).

\begin{figure}
\begin{center}
\includegraphics[width=3.6in, height=2.0in,clip=]{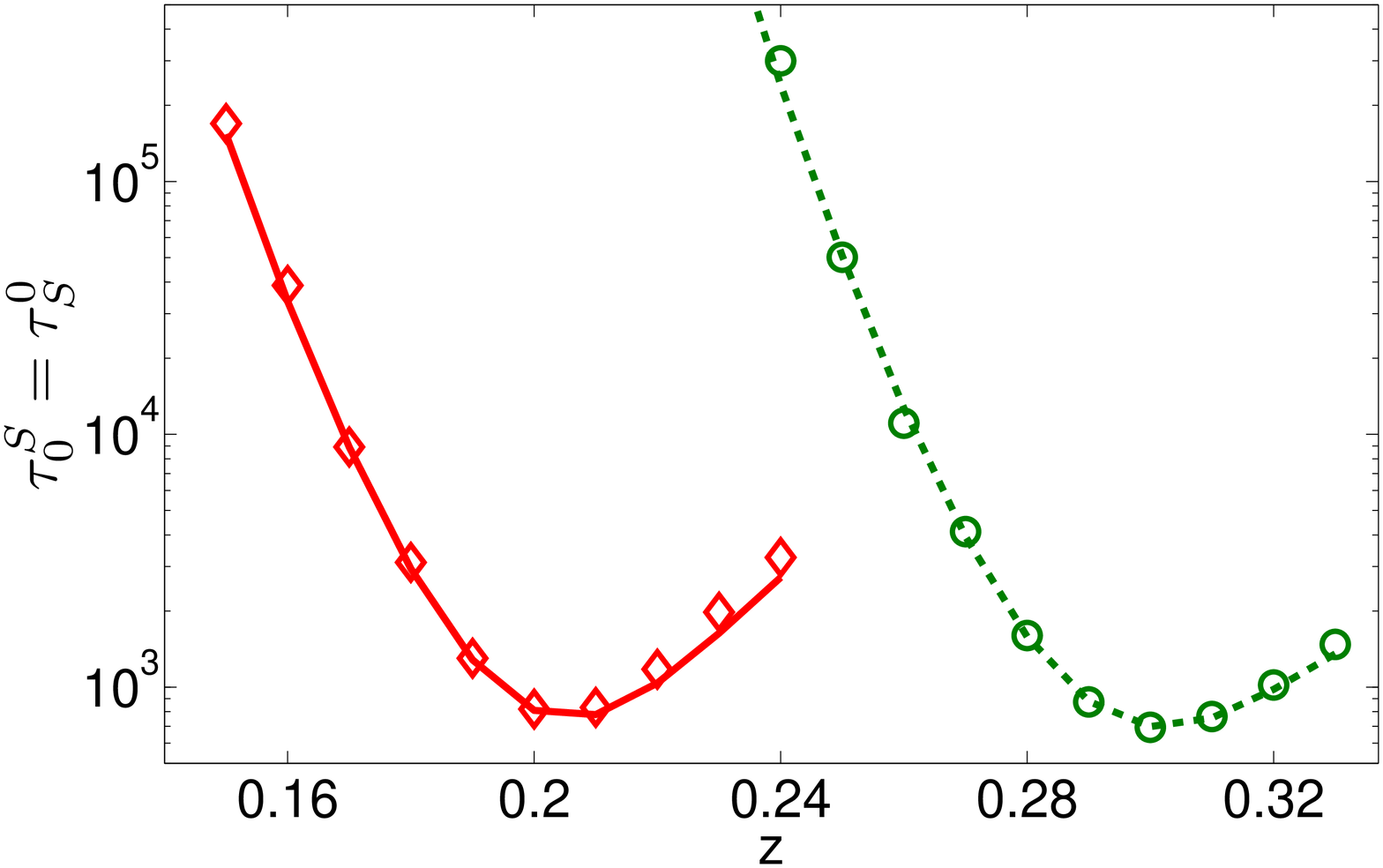}
\caption{(\textit{Color Online}) 
Mean switching time $\tau_0^S$ vs. $z$ in the symmetric case $z_+=z_-=z<z_c$ for
$q=2$ ($\Diamond$, solid), $q=3$ ($\circ$, dashed) and $N=100$. Symbols are  results of stochastic simulations and lines are the predictions of  (\ref{TOS1}), see text. 
}
\label{Fig8}
\end{center}
\end{figure}

The predictions of (\ref{TOS1})  are  reported in Fig.~\ref{Fig8} 
for various values of $z<z_c$. These are in good agreement   with the results of numerical simulations (averaged over 1000 samples, each run for $10^6$ simulation steps).
 When  $z$ is lowered well below $z_c$, the peaks of the SPD approach $n=0$ and $n=S$. In this case,
$\tau_0^S$  increases and switching allegiance takes very long. At fixed $z<z_c$, we find that $\tau_0^S$ increases with $q$.
 Interestingly, we also find that $\tau_0^S$ can exhibit a non-monotonic dependence on $z$ just below $z_c$ when
 $q$ is kept fixed, as shown in Fig.~\ref{Fig8}.

\subsection{Time-scale separation and growing fluctuations in the asymmetric case}
In the asymmetric case $0<z_-<z_+$, the party A has more zealot supporters than  party B. In this situation, when $z<z_c$ the SPD has a marked peaked near  $n=S$, see Fig.~\ref{Fig6}(a). As shown in Fig.~\ref{sample_fig}(b), the long time dynamics is characterized by a large majority of susceptibles becoming A supporters independently of the initial state. The magnetization $m(t)=2\delta z + [2n(t)-S]/N$ thus fluctuates around its MF value $m_+^{*}$ before reaching $m=m_{{\rm max}}=s+2\delta z$ when all susceptibles are supporters of party A, see Fig.~\ref{Fig9}(a). The population composition then endlessly fluctuates, with a majority of susceptibles supporting party A. In this case, with Eq.~(\ref{QmS}), the stationary ensemble-averaged magnetization
  $\langle m(\infty)\rangle =\sum_{m=m_{{\rm min}}}^{m_{{\rm max}}} m Q_m^*$ is positive.

 The $q$VM dynamics is thus characterized by  various regimes not captured by the mean field description. For concreteness, we consider that the initial density of A-susceptibles is $x_0<x^*$, as in Fig.~\ref{Fig9}, and distinguish four time scales: 

\begin{figure}
\begin{center}
\includegraphics[width=3.6in, height=2.0in,clip=]{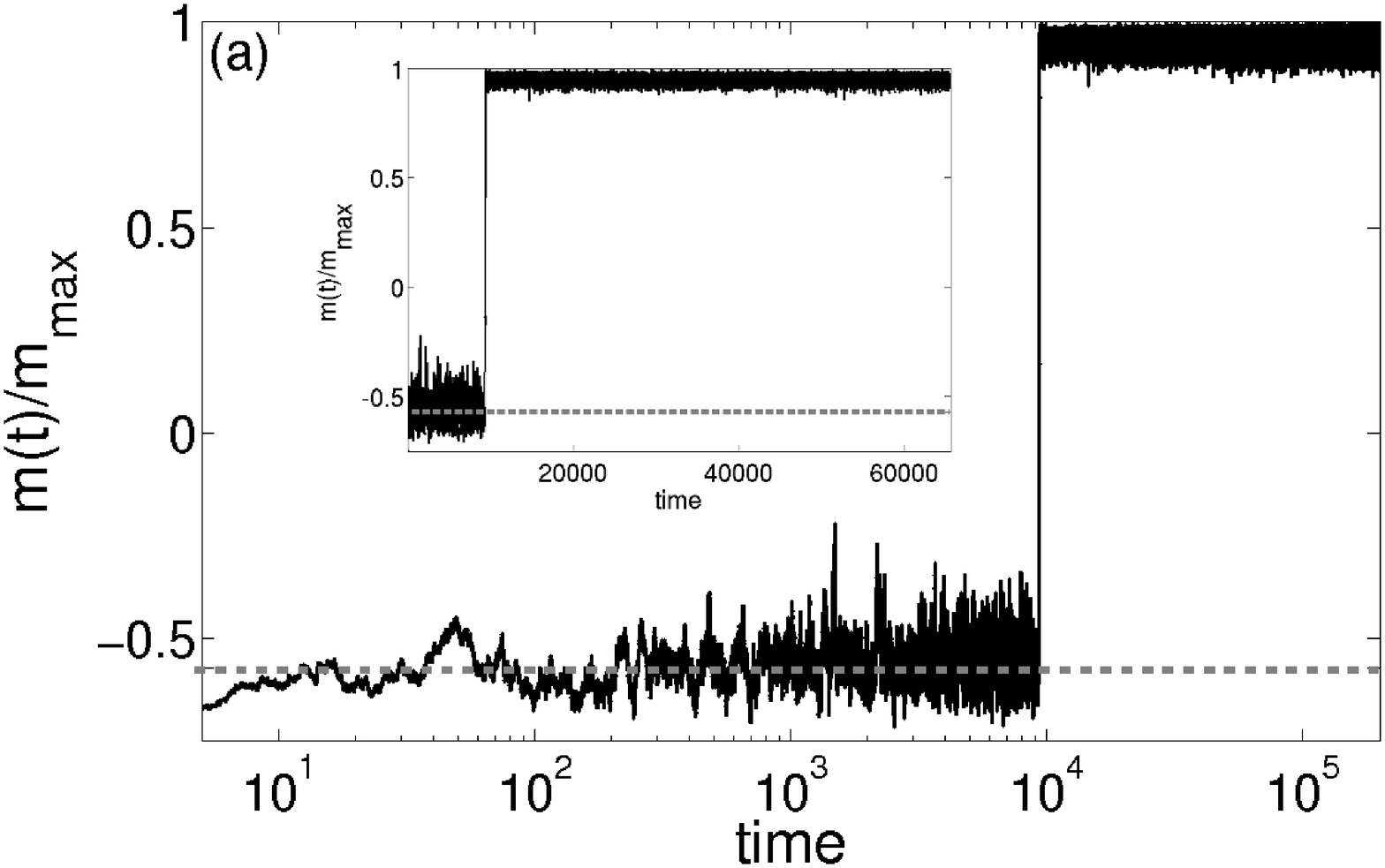}
\includegraphics[width=3.6in, height=2.0in,clip=]{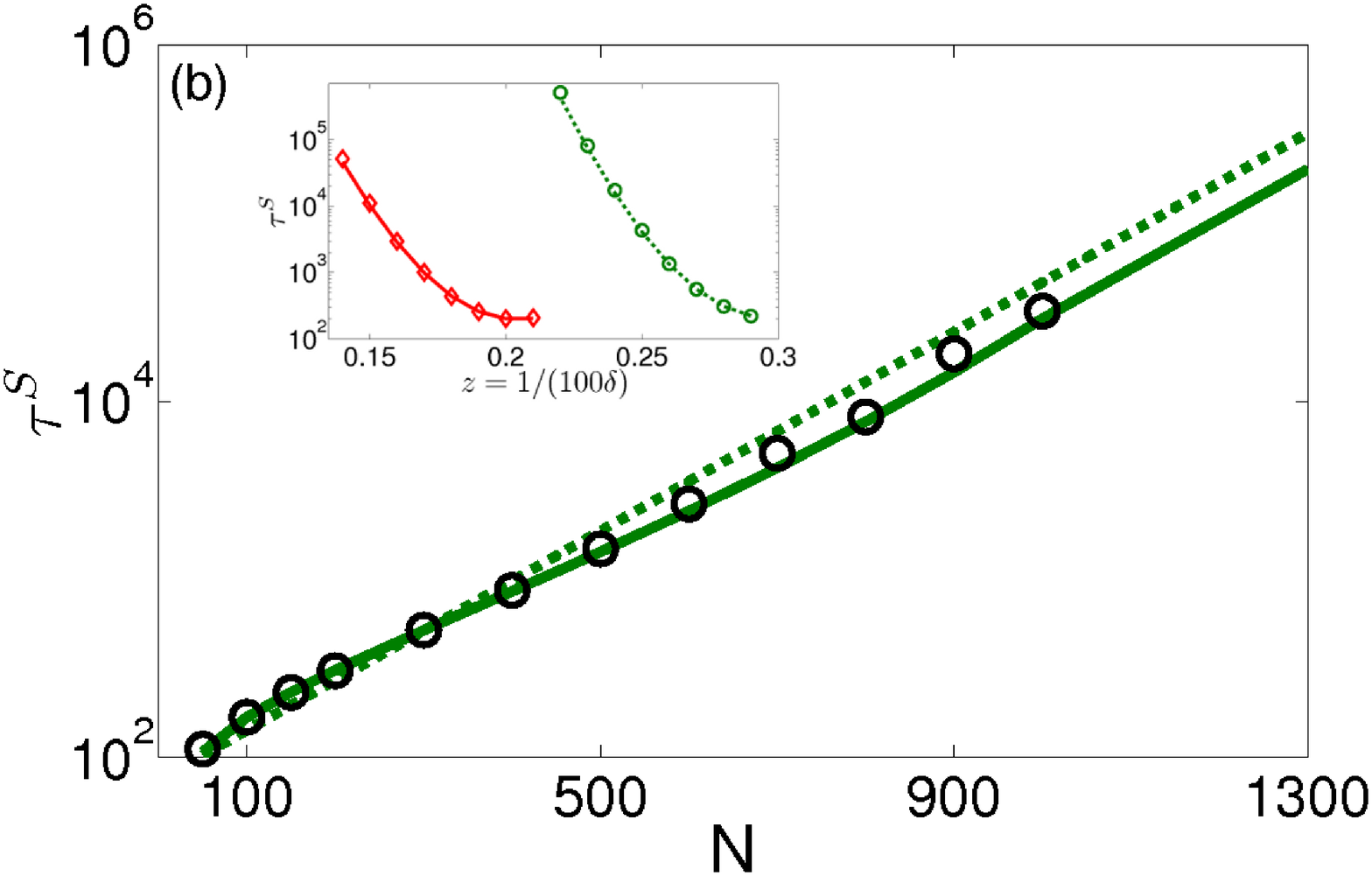}
\caption{(\textit{Color Online}) 
(a) Typical single realization of
 $m/m_{{\rm max}}=m/(s+2\delta z)$ in the asymmetric case at low zealotry, with
  $q=3$ and $m(0)=m_{{\rm min}}=-s+2\delta z$ 
(all B-susceptibles).
Here, $z=0.25$, $\delta=0.12$ and $N=1000$. 
After a time $t\approx 15$ the magnetization fluctuates around $m_{-}^*$ (dashed line); at time
  $t\approx  6\cdot 10^4$  it attains
  $m_{{\rm max}}=s+2\delta z$. In principle,
the magnetization can return to  $m_{{\rm min}}$ (all B-susceptibles)
after an enormous time ($\sim 10^{85}$, not shown).
Inset: Same, but in linear scale.
(b) $\tau^S$ as a function of $N$ for $q=3, z=0.25$ and $\delta=0.12$ (with $m(0)=m_{{\rm min}}$). 
Symbols ($\circ$) are the results of stochastic simulations. The solid and dashed 
lines are the predictions of
(\ref{TOS2}) and  (\ref{Tswitching}), respectively,  showing that
$\tau^S$ grows approximately exponentially with $N$, see text. 
Inset:  $\tau^S$ vs.
$z=(100\delta)^{-1}$ with $N=100$, and $Z_{+} - Z_{-}=2$ kept fixed,
 for $q=2$ ($\diamond$, solid) and $q=3$ ($\circ$, dashed). 
}
\label{Fig9}
\end{center}
\end{figure}

(i) After a mean time of order $\tau_{r_1}$, the system quickly relaxes toward
the metastable state $n_{-}^*$ where a random voter has the
MF opinion $m_{-}^*$, see Fig.~\ref{Fig9}(a).

(ii) After a mean time $\tau_{-}^{+}$, almost all  realizations suddenly approach the metastable state 
$n_{+}^*$  where $m(t)\approx m_{+}^*$, see Figs.~\ref{sample_fig}(a) and \ref{Fig9}(b).
The mean transition time $\tau_{-}^{+}$, as well as the average relaxation times,
can be estimated using Kramers' classical escape rate theory~\cite{Kramers}. The latter gives
the mean transition time $\tau_{{\rm K}}$ between the two local minima of the double-well potential $U(x)$ in which  an overdamped Brownian particle is
moving subject to a zero-mean delta-correlated Gaussian white noise force $\xi(t)$.
Here, we consider a potential $U(x)$ such that  $dU/dx=T^-(x)-T^+(x)$,
and the  noise correlations $\langle \xi(t) 
\xi(t')\rangle=\delta(t-t')~[T^+(x_{-}^*)+T^-(x_{-}^*)]/N$.
The bFPE generator of this Brownian particle is (\ref{G}) 
with a constant diffusive term evaluated at $x_{-}^*$. Kramer's formula hence gives~\cite{Kramers,KramersReview}:
\begin{eqnarray*}
\label{TK}
\tau_{-}^{+}\simeq \tau_{{\rm K}}= 2\pi~ \tau_{r_1}\tau_{r_2}~e^{2N\int_{x_{-}^*}^{x^*}
\frac{T^-(y)-T^+(y)}{T^-(x_{-}^*)+T^+(x_{-}^*)}~dy},
\end{eqnarray*}
where $\tau_{r_1} = 1/\sqrt{U''(x_{-}^*)}$, and $\tau_{r_2} = 1/\sqrt{|U''(x^*)|}$
denotes the mean relaxation time
from  state
$n=n^*$ to $n=n_{+}^*$.

(iii) The system then fluctuates around   $n_{+}^*$ 
before  reaching the state $n=S$ (all A-susceptibles) where $m=m_{{\rm max}}$, see Fig.~\ref{Fig9}(a) after a mean time $\tau^S$. In the realm of the bFPE, the mean time $\tau^S$ for all susceptibles to become A supporters for the first time
is
\begin{eqnarray}
\label{TOS2}
\tau^S= 2N\int_{x_0}^{s} dy~e^{-N\phi(y)}~\int_0^{y} \frac{e^{N\phi(v)}~dv}{T^+(v) + T^- (v)}.
\end{eqnarray}
When $n_+^*$ is close to the state $n=S$, the main contribution to $\tau^S$ is given by the mean transition time $\tau_{-}^{+}$ that is independent of $x_0<x^*$,  as illustrated by Fig.~\ref{Fig9}(b). This is well approximated by Kramer's formula, yielding
\begin{eqnarray}
\label{Tswitching}
\tau^S&\sim& \tau_{-}^{+}\simeq 2\pi~ \tau_{r_1}\tau_{r_2}~e^{2N\int_{x_{-}^*}^{x^*}
\frac{T^-(y)-T^+(y)}{T^-(x_{-}^*)+T^+(x_{-}^*)}~dy},
\end{eqnarray}
showing that the mean switching time scales exponentially with the population size.

(iv) The amplitude of the fluctuations around $n\approx S$, where $m(t)\approx m_{{\rm max}}$ grows endlessly in time, see Fig.~\ref{sample_fig}(b), and  the system eventually returns to the state $n=0$ (all B-susceptibles). Yet, this occurs after an enormous amount of time, of order $e^{2N\int_{x_{+}^*}^{x^*}
\frac{T^-(y)-T^+(y)}{T^-(x_{+}^*)+T^+(x_{+}^*)}~dy}$,  that is generally not physically observable when $N\gg 1$.

The predictions (\ref{TOS2}) and its approximation (\ref{Tswitching}) are reported in Fig.~\ref{Fig9}(b), where they are in good agreement with the results of stochastic simulations. These results confirm that $\tau^S$ grows approximately exponentially with $N$ when $N\gg 1$.
In Fig.~\ref{Fig9}(b), we also see that   $\tau^S$ increases with $1/z$, 
and with $q$ when $z$ and $\delta$ are fixed. As illustrated in Fig.~\ref{Fig9}(a), contrary to the case of symmetric zealotry, there is no ``swing-state dynamics": After a mean time $\tau^S$ the population persists near $n\approx S$ where most susceptibles are A supporters and the magnetization is 
$m\approx m_{{\rm max}}$, and there is virtually no switching back to state  $n\approx 0$. Hence,  a small bias in the zealotry, combined with  
fluctuations and nonlinearity, can greatly affect the voters' opinion in the $q$VM.

\section{Mean consensus time in the presence of one type of zealots}
When there are only A-zealots,  with   $z_+=\zeta$  and $z_{-}=0$, an A-party consensus is always reached. Yet, the dynamics leading to the corresponding absorbing state $n=S$ depends non-trivially on the  zealotry density and on the initial density $x_0$ of A-susceptibles. 

Here, the  fluctuation-driven dynamics is characterized by the mean  consensus time (MCT). As illustrated in Fig.~\ref{Fig10}, the MCT can change by several order of magnitudes when $\zeta$ and $x_0$ change over a small range:
(i) Below the critical zealotry density $\zeta_c$, the MCT grows exponentially with 
the population size $N$ when  $x_0<x^*$, see Fig.~\ref{Fig10}(b); (ii) Otherwise the MCT grows logarithmically with $N$, see Fig.~\ref{Fig10}(inset).
These  phenomena are analyzed as follows: 
\begin{figure}
\begin{center}
\includegraphics[width=3.6in, height=2.0in,clip=]{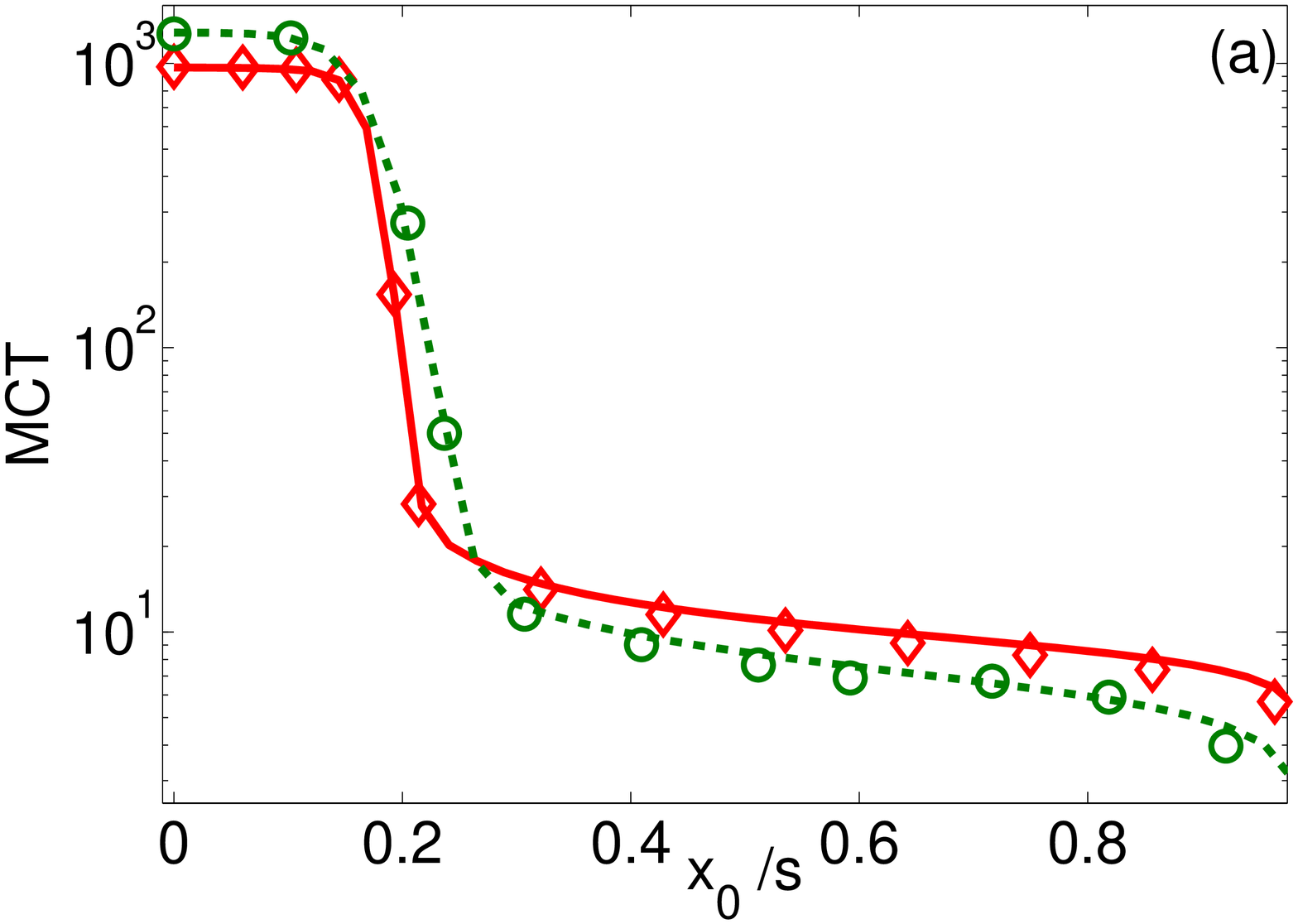}
\includegraphics[width=3.6in, height=2.0in,clip=]{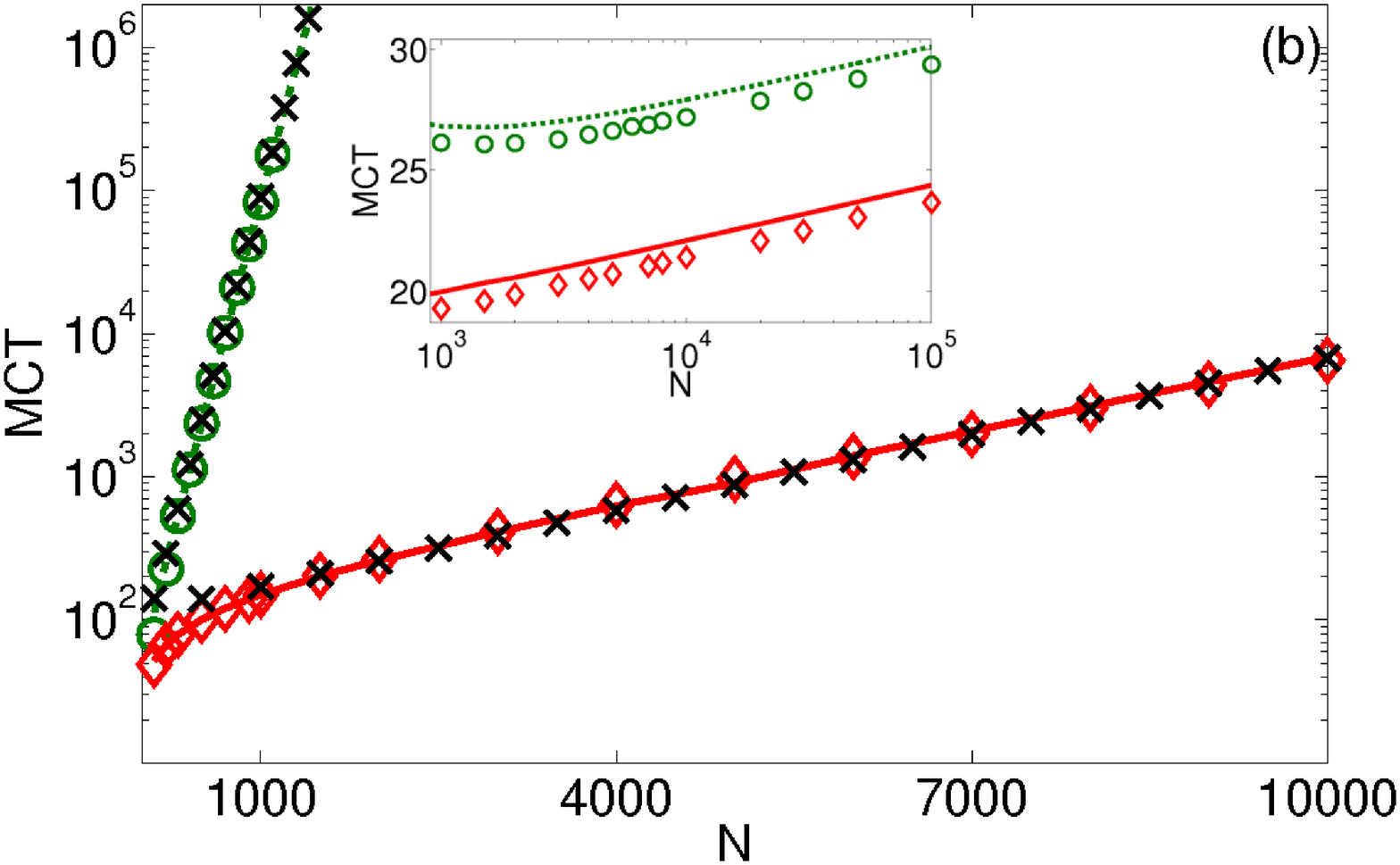}
\caption{(\textit{Color Online}) 
(a) MCT vs. $x_0/s$ for
$q=2, \zeta=0.17$ and $N=5000$ ($\diamond$, solid), and $q=3, \zeta=0.24$ and $N=400$ ($\circ$, dashed).
Symbols are from stochastic simulations and  lines are the predictions of (\ref{MCTdiff}).
(b) MCT vs. $N$ for
$q=2, \zeta=0.17$ ($\diamond$, solid), and for $q=3, \zeta=0.24$  ($\circ$, dashed).
Initially $x_0=0.1< x^*$.
Symbols ($\diamond$) and ($\circ$) are from  simulations (averaged over $10^3-10^5$ samples), lines are the predictions of (\ref{MCTdiff})
and ($\times$) are proportional to the WKB results, Eqs. (\ref{MCTWKB}) and (\ref{Sq2q3}).
Inset:  MCT vs. $N$ in semi-log scale for
$q=2, \zeta=0.2$ ($\diamond$, solid), and for $q=3, \zeta=0.26$ and ($\circ$, dashed).
}
\label{Fig10}
\end{center}
\end{figure}

(i) When $\zeta\leq \zeta_c$ and $x_0<x^*$, in line with the MF analysis, the density of A-susceptibles  first lingers around the metastable state $x_b^*$  until a large fluctuation drives the system towards the absorbing state. This large-fluctuation-driven phenomenon is particularly well captured by the WKB theory~\cite{WKB,otherWKB,3SVMZ}. The essence of this method consists of studying the quasi-stationary probability distribution (QSPD) $\pi_n$ obtained  by setting
$P_{n}(t)\simeq \pi_{n}~e^{-t/\tau_c} $ for $0\leq n<S$
and $P_S(t)\simeq 1-e^{-t/\tau_c}$ into the master equation (\ref{ME}). The MCT is the mean decay time $\tau_c$ of the QSPD. Since $(d/dt)P_S\simeq e^{-t/\tau_c}/\tau_c\simeq T^+_{S-1}\pi_{S-1}e^{-t/\tau_c}$, we indeed find~\cite{WKB,otherWKB}
\begin{eqnarray}
\label{tauC}
\tau_c=(T^+_{S-1}\pi_{S-1})^{-1}.
\end{eqnarray}
The computation of the MCT therefore requires finding the  QSPD. This obeys
\begin{eqnarray}
\label{QSD}
 T^{+}_{n-1} \pi_{n-1} + T^{-}_{n+1} \pi_{n+1} - (T_n^+ + T_n^-)\pi_n=0,
\end{eqnarray}
obtained from Eq.~(\ref{ME}) upon neglecting an exponentially small term $\pi_n/\tau_c$. 
In the limit $N\gg 1$, the density $x=n/N$ is treated as a continuous variable and
Eq.~(\ref{tauC})  yields 
$\tau_c^{-1}= (\pi(s)/N)
\left|\frac{d}{dx}~T^+(x)\right|_{x=s}$. In the continuum limit,  Eq.~(\ref{QSD}) is solved with the WKB Ansatz
\begin{eqnarray}
\label{WKB}
\pi(x)\simeq {\cal A}~e^{-N{\cal S}(x) -{\cal S}_1(x)},
\end{eqnarray}
where ${\cal S}(x)$ is the action, ${\cal S}_1(x)$ is the amplitude, and $ {\cal A}\sim e^{N {\cal S}(x_b^*)}$ is a normalization constant
~\cite{otherWKB}. By substituting (\ref{WKB}) into (\ref{QSD}),
to leading order we find~\cite{WKB,otherWKB}
\begin{eqnarray}
\label{action}
{\cal S}(x)=-\int^{x} \Psi(y)~dy,
\end{eqnarray}
where, as in Sec.~IV, $\Psi(y)=\ln{[T^{+}(y)/T^{-}(y)]}$.

 Hence, when $x_0<x^*$ and $N\gg 1$, the leading contribution to the MCT  is given by the accumulated action $\Delta {\cal S}$ over the path joining the metastable state $x=x_b^*$ and the unstable steady state $x=x^*$~\cite{WKB,otherWKB}:
\begin{eqnarray}
\label{MCTWKB}
\tau_c\sim e^{N[{\cal S}(x^*)-{\cal S}(x_b^{*})]}=e^{N\Delta {\cal S}}.
\end{eqnarray}
The next-to-leading correction arising from ${\cal S}_1(x)$ is given in Refs.~\cite{WKB,otherWKB},
but for our purpose Eq.~(\ref{MCTWKB}) already provides  useful information on the MCT. In fact, for $q=2$ and $q=3$, the action (\ref{action}) explicitly reads
\begin{eqnarray}
\label{Sq2q3}
-{\cal S}(x)=\left\{
  \begin{array}{l l } 
&(1-\zeta)\ln{(1-\zeta-x)}+2\zeta\ln{(x+\zeta)}\\
&+x\ln{\left[
\frac{(x+\zeta)^2}{x(1-\zeta-x)}
\right]} 
\quad \hspace{10.16mm} \text{($q=2$)}\\
 &   \zeta\ln{x}+ 2\ln{(1-\zeta-x)}\\
&+(x+\zeta)\ln{\left[
\frac{(x+\zeta)^3}{x(1-\zeta-x)^2}\right]} 
 \quad \text{($q=3$)}\\
  \end{array}\right.
\end{eqnarray}
With these expressions, and with  (\ref{xb}) and (\ref{xu}) for  $x_{b}^*$ and $x^*$,
the leading  contribution to the WKB approximation of the MCT  is computed explicitly, and the results reported in Fig~\ref{Fig10}(b) are in excellent agreement with those of stochastic simulations when $N\gg 1$ and confirm that $\tau_c$ grows exponentially with $N$. We can also check that $\Delta {\cal S}$ is a decreasing function of $\zeta$, which clearly implies that the MCT grows when $\zeta$ is decreased.

\vspace{0.1cm}

(ii) When $\zeta>\zeta_c$, or for any $\zeta>0$ when the initial density $x_0>x^*$, the A-party  consensus is reached much quicker than in the case (i), typically after a time of order ${\cal O}(\ln{N})$, see Fig.~\ref{Fig10}(inset). The backward Fokker-Planck formalism is again suitable to derive this result. In such a framework, the MCT obeys 
Eq.~(\ref{bFPE})  supplemented by reflective and absorbing boundary conditions $\tau_c'(0)=0$ and $\tau(s)=0$~\cite{noise}. Proceeding as in Sec.~V, we find again the  expression:
\begin{eqnarray}
\label{MCTdiff}
\hspace{-5mm}
\tau_c(x_0)= 2N\int_{x_0}^{s} dy~e^{-N\phi(y)}~\int_0^{y} \frac{e^{N\phi(v)}~dv}{T^+(v) + T^- (v)},
\end{eqnarray}
with $\phi(v)=-2\int_0^v du~\left\{\frac{T^-(u) - T^+ (u)}{T^-(u) + T^+ (u)}\right\}$.
As shown in the inset of Fig.~\ref{Fig10}, this expression is in good agreement with the results of stochastic simulations and captures the functional dependence of the MCT
whose  leading contribution grows logarithmically with $N$ and increases with $q$. 
It is also worth noting that Eq.~(\ref{MCTdiff}) also provides a meaningful approximation of the MCT 
in the metastable regime, even though when $N\gg 1$ its predictions are usually less accurate  than those of the WKB method, see {\it e.g.}~\cite{otherWKB,3SVMZ}.

\section{Summary and conclusion}
We have studied the dynamics of the non-linear $q$-voter model ($q$VM) in the presence of 
inflexible zealots in a finite well-mixed population. In this model, voters can support two parties and 
are  either  ``susceptibles'' or ``inflexible zealots''. Susceptible voters adopt the opinion of a group of $q\geq 2$ neighbors if they all agree, while zealots are here individuals whose state never changes.
The $q$VM with zealots is introduced  as a  simple non-trivial model able to capture the essence of important concepts of social psychology and sociology, such as the relevance of conformity and independence as mechanisms for collective  actions~\cite{ConfIndep,Granovetter}, and the existence of group-size threshold that  influences the social impact of conformity~\cite{GroupSize}.

In spite of its simplicity and the fact that the detailed balance is satisfied, the dynamics of the non-linear $q$VM with zealots is rich and characterized by fluctuation-driven phenomena and  non-trivial probability distributions. The dynamics is particularly interesting at low level of zealotry, when the stationary distribution is bimodal. In this case, we have found that when one party has more zealots than the other,
 the  intensity of one peak greatly exceeds that of the other. The dynamics is thus characterized by various time scales and growing fluctuations around a state in which a majority of susceptibles support the party having more zealots. 
When both parties have the same number of zealots, below the critical zealotry, the long-time dynamics is characterized by the susceptibles endlessly swinging from a state in which they all support one party to the state where they all support the other party. We have rationalized all these features 
by computing the exact stationary probability distribution and, within the backward Fokker-Planck formalism, the mean times for all susceptibles to switch allegiance. We have thus found that these mean switching times grow approximately exponentially with the population size, and they increase when the number of zealot decreases  at low zealotry. When zealots support only one party, we have shown that a consensus is reached in a mean time that grows either exponentially or logarithmically with the population size, depending on the zealotry density and the initial condition.
\\
Our findings show that the properties of the nonlinear $q$VM  with zealots ($q\geq 2$) are dominated by fluctuations, and have revealed that they are sensitive  to even a small bias in the zealot densities. Most of the features of the nonlinear $q$VM  with inflexible zealots are therefore beyond the reach of  a simple mean field analysis and generally deviate from those of the classical linear voter model.


\begin{thebibliography}{99}
\bibitem{Comment}
In the formulation of Ref.~\cite{qVM}, 
the agent can still change its opinion with a probability $\epsilon$, even if not all its $q$ neighbors agree. For simplicity,  here we have $\epsilon=0$ as in \cite{exitprobq}.
\end{thebibliography}
\end{document}